\documentclass[prd,aps,twocolumn,a4paper,floatfix]{revtex4}

\usepackage{graphicx,psfrag}
\usepackage{mathrsfs}
\usepackage{amsmath,amsfonts,amssymb}
\usepackage{multirow}
\usepackage{comment}

\newcommand{\be}{\begin{equation}}
\newcommand{\ee}{\end{equation}}
\newcommand{\bea}{\begin{eqnarray}}
\newcommand{\eea}{\end{eqnarray}}
\newcommand{\bel}{\begin{align}}
\newcommand{\eel}{\end{align}}

\newcommand{\avg}[1]{\left< #1 \right>}

\def\non{\nonumber}                     

\def\bam{{\textsc{BAM}}}
\def\lorene{{\textsc{Lorene}}}

\def\Msun{M_{\odot}}
\def\GMc2{G M_{\odot} c^{-2}}
\def\Mpc{Mpc}

\def\sf{{\rm SF}}

\def\I{\mathcal{I}}
\def\M{\mathcal{M}}
\def\O{\mathcal{O}}
\def\vareps{\varepsilon}
\def\vrho{\varrho}
\def\rad{rad}

\usepackage{color}

\definecolor{cyan}{rgb}{0,0.9,0.9}
\definecolor{orange}{rgb}{0.9,0.5,0}
\definecolor{magenta}{rgb}{1,0,1}
\definecolor{purple}{rgb}{0.8,0.4,0.8}
\definecolor{gray}{rgb}{0.8242,0.8242,0.8242}

\begin{document}

\title{Accuracy of numerical relativity waveforms from binary neutron
  star mergers and their comparison with post-Newtonian waveforms}

\author{Sebastiano \surname{Bernuzzi}}

\author{Marcus \surname{Thierfelder}}

\author{Bernd \surname{Br\"ugmann}}

\affiliation{Theoretical Physics Institute, University of 
  Jena, 07743 Jena, Germany}

\date{\today}

\begin{abstract}
  We present numerical relativity simulations of nine-orbit 
  equal-mass binary neutron star covering the quasicircular late
  inspiral and merger. 
  The extracted gravitational waveforms are analyzed for convergence
  and accuracy. Second order convergence is observed up to contact,
  i.e.~about 3-4 cycles to merger; error estimates can be made up to
  this point.
  The uncertainties on the phase and the amplitude are dominated by
  truncation errors and can be minimized to $0.13$~$\rad$   
  and $\lesssim1~\%$, respectively, by using several simulations
  and extrapolating in resolution. 
  In the latter case finite-radius extraction uncertainties become 
  a source of error of the same order and have to be
  taken into account.
  The waveforms are tested against accuracy standards for data analysis.
  The uncertainties on the waveforms are such that accuracy standards 
  are generically not met for signal-to-noise ratios relevant for
  detection, except for some best cases using extrapolation from several runs. 
  A detailed analysis of the errors is thus imperative for the
  use of numerical relativity waveforms from binary neutron stars
  in quantitative studies.
  The waveforms are compared with the post-Newtonian Taylor T4
  approximants both for point-particle and including the analytically
  known tidal corrections. 
  The T4 approximants accumulate significant phase differences of 
  $2$~$\rad$ at contact and $4$~$\rad$ at merger, 
  underestimating the influence of finite size effects.
  Tidal signatures in the waveforms are thus important at least during the last six orbits of the merger process.
\end{abstract}


\maketitle

\section{Introduction}
\label{sec:intro}

An exciting possibility to reveal and to study the nature of the
neutron star (NS) interior is provided by the detection of
gravitational waves (GWs) from binary neutron star (BNS) mergers. 
The GWs emitted by BNS during the late inspiral and merger are sensitive to
finite size effects, and in particular to the tidal interaction between
the bodies, thus they are quantitatively dependent on the star parameters
and, in turn, on the equation of state (EoS).
Ground-based interferometers are sensitive to the last $10$ 
orbits of a typical equal-mass binary system of mass $\sim2.8~\Msun$, 
which roughly corresponds to the frequency range $400-1500~Hz$.
During this phase tidal effects are expected to be significant.

A general relativistic perturbative theory of
tidal interactions has been developed 
in recent
years~\cite{Hinderer:2007mb,Damour:2009vw,Binnington:2009bb}.    
These results have been incorporated into the post-Newtonian~(PN) 
formalism~\cite{Hinderer:2009ca,Vines:2010ca,Vines:2011ud}, thus
permitting the extension of phasing formulas to tidally interacting
binaries, as well as into the effective-one-body (EOB) 
model~\cite{Damour:2009wj}. 

An exact and quantitative evaluation of the dynamics and of the
waveforms during the merger process requires, however, the solution of
the full nonlinear Einstein equations. In particular, numerical
relativity~(NR) simulations are to date the only tool to tackle the
problem (see
e.g.~\cite{Giacomazzo:2010bx,Baiotti:2011am,Hotokezaka:2011dh,Sekiguchi:2011zd,Shibata:2011fj} 
for recent works in the field
and~\cite{Faber:2009,Duez:2009yz,Rosswog:2010ig} for reviews). 

Numerical relativity data have been used in combination with 
PN methods in order to assess the detectability of tidal
effects and the accuracy of the parameter estimated from GW 
measurements~\cite{Read:2009yp}. 
A more recent application of NR results concern their use for
calibrating the tidal-EOB model~\cite{Baiotti:2010xh,Baiotti:2011am}.
These works highlight the importance of using NR waveforms and 
analytic results in order to quantitatively evaluate the
impact of tidal effects in the waveforms.

An aspect which deserves a more detailed assessment
than is currently available in the literature is the accuracy of the
NR waveforms.
The estimates of error-bars on phase and amplitude of BNS waveforms, as
well as their assessment against accuracy standard for
detection~\cite{Miller:2005qu,Lindblom:2008cm,Lindblom:2009ux}, 
is of fundamental importance for any quantitative study.
In constrast to binary black holes (BBHs) data, whose quality for data
analysis purposes is well documented, see 
e.g.~\cite{Hannam:2009hh,Reisswig:2009vc,Hannam:2010ky,MacDonald:2011ne},
convergence and uncertainties in BNS simulations are so far poorly investigated. 
To our knowledge, the only analysis of truncation errors have been performed
in~\cite{Baiotti:2009gk}, and, more
exhaustively, in~\cite{Thierfelder:2011yi}. Both works found that the waveforms 
are second order convergent up to merger  but they are limited to
short runs (three orbits) and do not consider accuracy criteria for detection.
In~\cite{Baiotti:2011am} the same initial data as in this work have been evolved, 
error bars considering finite-extraction effects and truncation errors have been 
estimated but without performing convergence tests and using only two simulations. 
In their conclusions the authors stressed the need for a detailed error 
budget based on convergence measurements.  
Since the numerical treatment of the matter (hydrodynamics) makes very
challenging to obtain accurate waveforms (in comparison with BBHs
simulations), a precise and rigorous assessment of their quality is urgent.

In this work we report results about the accuracy of the 
waveforms extracted from BNS simulations. Focusing on an example
configuration, we consider nine-orbit simulations employing different
resolutions, and reaching the highest resolutions for production
runs used so far.
We discuss convergence in detail, compute error-bars of the waveform amplitude
and phase, and test them against accuracy standards for detection.
The NR waveforms are then contrasted with the post-Newtonian (PN) T4
phasing formula, both for point-particle and including the
leading-order and next-to-leading-order tidal corrections analytically
known. 

The structure of the paper is as follows. 
In Sec.~\ref{sec:framework} we review our methodology.
In Sec.~\ref{sec:dynamics} the simulated binary dynamics is presented.
In Sec.~\ref{sec:waves} we analyze the waveforms.
In particular, convergence is analyzed in 
Sec.~\ref{sbsec:conv}, 
the influence of finite-radius extraction is analyzed in
Sec.~\ref{sbsec:finiter}, and 
the waveforms are tested against accuracy standards in 
Sec.~\ref{sbsec:acc}.  
In Sec.~\ref{sec:comparepn} the numerical waves are compared with the
analytic post-Newtonian T4 formula including tidal effects. 

Dimensionless units $c$=$G$=$\Msun$=1 are used in this paper, unless
otherwise stated.

\section{Theoretical and numerical setup}
\label{sec:framework}

In this section we outline the theoretical and numerical frameworks
employed in this paper, and we describe the setup of the simulations.
More details on the methodology are given
in~\cite{Brugmann:2008zz,Thierfelder:2011yi} and references therein.

The present study relies on evolutions of BNS initial data within 3+1
numerical relativity, using the BSSNOK 
formulation~\cite{Nakamura:1987zz,Shibata:1995we,Baumgarte:1998te} of
Einstein equations coupled with the general relativistic hydrodynamics
(GRHD) system~\cite{Banyuls:1997zz}. 
The gauge is specified by the 1+log lapse and
Gamma-driver-shift~\cite{Bona:1994b,Alcubierre:2002kk,vanMeter:2006vi,Gundlach:2006tw}, 
using the same expressions and parameters of Sec.~II
of~\cite{Thierfelder:2011yi}. 
In particular the damping parameter for the shift equation is set to $\eta=0.3$.
Gravitational waves are extracted from the numerically generated
spacetime by using the Newman-Penrose scalar, $\psi^4$. The projections
onto spin weighted spherical harmonics, i.e.~the multipoles of the
radiation, are evaluated on extraction sphere of finite coordinate radius $r$.
The same conventions of~\cite{Brugmann:2008zz} are employed here.
The diagnostic quantities discussed in the following are the ADM mass,
$M_{\rm ADM}$, the Hamiltonian constraint, $Ham$, and the rest-mass
integral, $M_0$, Cf.~Eq.~(31) of~\cite{Thierfelder:2011yi}.

The code employed in this work is the {\bam}
code~\cite{Thierfelder:2011yi,Brugmann:2008zz,Bruegmann:2003aw,Bruegmann:1997uc},
which implements finite differencing methods on Cartesian
refined meshes. 
The evolution algorithm is based on the method of lines and explicit
Runge-Kutta methods (third order in this work).
A combination of centered and lop-sided standard finite differences in
space is used for the metric fields, see~\cite{Brugmann:2008zz}. 
In this work fourth order operators are employed, together with sixth
order artificial dissipation.
The algorithm implemented for the matter is a robust
high-resolution-shock-capturing scheme based on a central scheme for the 
numerical
fluxes~\cite{DelZanna:2007pk,Kurganov:2000,Nessyahu:1990,Shu:1988,Shu:1989}. Both 
the time stepping and the spatial refined mesh are shared with the metric system.  
The interface fluxes are computed by the local Lax-Friedrichs (LLF)
central scheme~\cite{Nessyahu:1990,Kurganov:2000}, while
reconstruction is performed with the third order 
convex-essentially-non-oscillatory (CENO) interpolation~\cite{Liu:1998,Zanna:2002qr}. 
Mesh refinement is provided by a hierarchy of cell-centered nested
Cartesian grids and Berger-Oliger time stepping. 
Metric variables are interpolated in space by means of fourth order 
Lagrangian polynomials and matter conservatives by a fourth order
weighted-essentially-non-oscillatory (WENO) scheme~\cite{Macdonald:2008}.
Interpolation in Berger-Oliger time stepping is performed at second order. 
Some of the mesh refinement levels can be dynamically moved and
adapted during the time evolution according to the technique of
``moving boxes'', e.g.~\cite{Brugmann:2008zz}. 

Initial data are chosen from quasi-equilibrium configurations of irrotational
equal-mass binaries in quasicircular orbits~\cite{Gourgoulhon:2000nn,Taniguchi:2002ns}.  
The configuration selected for this work is a binary with ADM mass
$M_{\rm ADM}=M=3.00506$, rest-mass $M_0=3.250$, and angular momentum
$J_{\rm ADM}=9.716$. 
The initial proper relative separation is
$d\simeq50$~($\sim70~km$) corresponding to a GWs frequency
$f_0=0.0019$~($394~Hz$).  
The compactness of each star in isolation is $0.14$. 
The EoS for the fluid is the polytropic one, with
adiabatic index $\Gamma=2$.
The initial configuration is computed with a multidomain spectral 
code which solves the Einstein constraint equations under the
assumption of a conformally flat metric. The code is based on the 
{\lorene} library~\cite{LORENE} and provided by the
NR group in LUTH~(Meudon). These initial 
data represent to date the most accurate computation of equilibrium
BNSs and they are publicly available on the web.
The same initial data were used for the evolutions discussed
in~\cite{Baiotti:2010xh,Baiotti:2011am}. 

\begin{table*}[t]
  \caption{ \label{tab:grid_runs} 
    Summary of the grid configurations and of the runs.
    Columns: 
    name of the configuration, 
    maximum refinement level, 
    minimum moving level, 
    number of points per direction in the moving levels, 
    resolution per direction in the level $l=l_{\rm max}$, 
    number of points per direction in the nonmoving levels, 
    resolution per direction in the level $l=0$, 
    number of processors, 
    maximal memory usage, and average speed in term of the mass of the
    configuration evolved ($M=2.998~\Msun$) 
    including checkpointing and initialization 
    (reference machine: JUROPA cluster).}  
  \centering    
  \begin{tabular}{ccccccccccc}        
    \hline
    Name & $l_{\rm max}$ & $l^{\rm mv}$ & $N_{\rm xyz}^{\rm mv}$ &
    $h_{\rm l_{\rm max}}$ & $N_{\rm xyz}$ & $h_0$ &
    Nproc & Mem (Gb) & Speed (M/hr)\\
    \hline
    HH2 & 7 & 4 & 100 & 0.1875 & 160 & 24    &  64 & 150 & 6\\
    HH3 & 7 & 4 & 128 & 0.1466 & 176 & 18.75 &  64 & 190 & 4.5\\
    HH4 & 7 & 4 & 140 & 0.1328 & 192 & 17.14 &  96 & 240 & 5\\
    HH5 & 7 & 4 & 150 & 0.1250 & 200 & 16    & 256 & 350 & 12\\
    HH6 & 7 & 4 & 160 & 0.1172 & 212 & 15    & 256 & 400 & 12\\
    \hline
  \end{tabular}
\end{table*}

Evolutions were performed for the $\Gamma=2$ polytropic EoS, 
i.e.~we consider the fluid isentropic and neglect thermal effects. 
In~\cite{Thierfelder:2011yi} we have shown that, in agreement with the
physical expectation, waveforms computed with both the
polytropic EoS and the ideal gas EoS (which includes in a rough way
thermal effects) are indistinguishable within the 
simulation errors, at least up to contact, while significant
differences accumulate during merger and the HMNS phase. 
In this work we are mainly interested in the inspiral phase, hence
thermal effects can be neglected; we will consider thermal effects in
the present setup in future work. 

Gravitational waves were extracted at levels $l=1,2,3$.
Several resolutions and a single grid setup were employed.
The latter is composed of a fundamental
grid level, $l=0$, and seven refinement levels from $l=1$ to $l_{\rm max}=7$;
four refinement levels are moving, $l=4,5,6,7$. 
The only symmetry assumed is reflection symmetry about the $z=0$ plane, i.e.~the
numerical domain is restricted to $z>0$.
The grid configurations, as well as the performances of the runs are
reported in Tab.~\ref{tab:grid_runs}. 
The grid settings are similar to those of other
codes, e.g.~\cite{Baiotti:2010ka}. 
The highest resolution employed here for run HH6 is slightly ($3\%$)
higher than the maximum resolution used to date on BNS simulations employing
mesh-refined-Cartesian-grid-based codes~\cite{Baiotti:2011am}.
All the runs were performed with Courant-Friedrich-Lewy (CFL) factor
of $0.25$.
The (self) convergent series are formed by triplets of runs, that in
the following will be denoted as HH$\{LMH\}$, where $L,M,H$ correspond to 
the low, medium, and high resolution employed.
Self-convergence tests can be biased by the choice of the resolutions
employed. An ``optimal'' setup would require that 
(i)~the ratios between the low and medium and medium and high
resolutions are $h_L/h_M\simeq h_M/h_H\simeq2$; and 
(ii)~the scaling factor is at least of order two,
$\sf=(h_L^r-h_M^r)/(h_M^r-h_H^r)\gtrsim2$, where $r$ is the convergence rate.
It is difficult to obtain an optimal convergent series since low resolutions are 
too inaccurate and differ even in a qualitative way~\cite{Thierfelder:2011yi}.
Considering the criteria above, the ``best'' convergent series
is HH$\{235\}$ \emph{or} HH$\{236\}$.

\section{Overview of the binary dynamics}
\label{sec:dynamics}

In this section we summarize the binary dynamics and present some
diagnostic of the simulations.

The binary evolves for about nine orbits dynamics before merger, when
a hyper-massive-neutron-star (HMNS) is formed. 
The latter oscillates non linearly in time, loses angular momentum by 
GW emission increasing its compactness, and, finally, collapses to a black
hole surrounded by a disk rapidly accreting. 

\begin{figure*}[t]
  \begin{center}
    \includegraphics[width=0.49\textwidth]{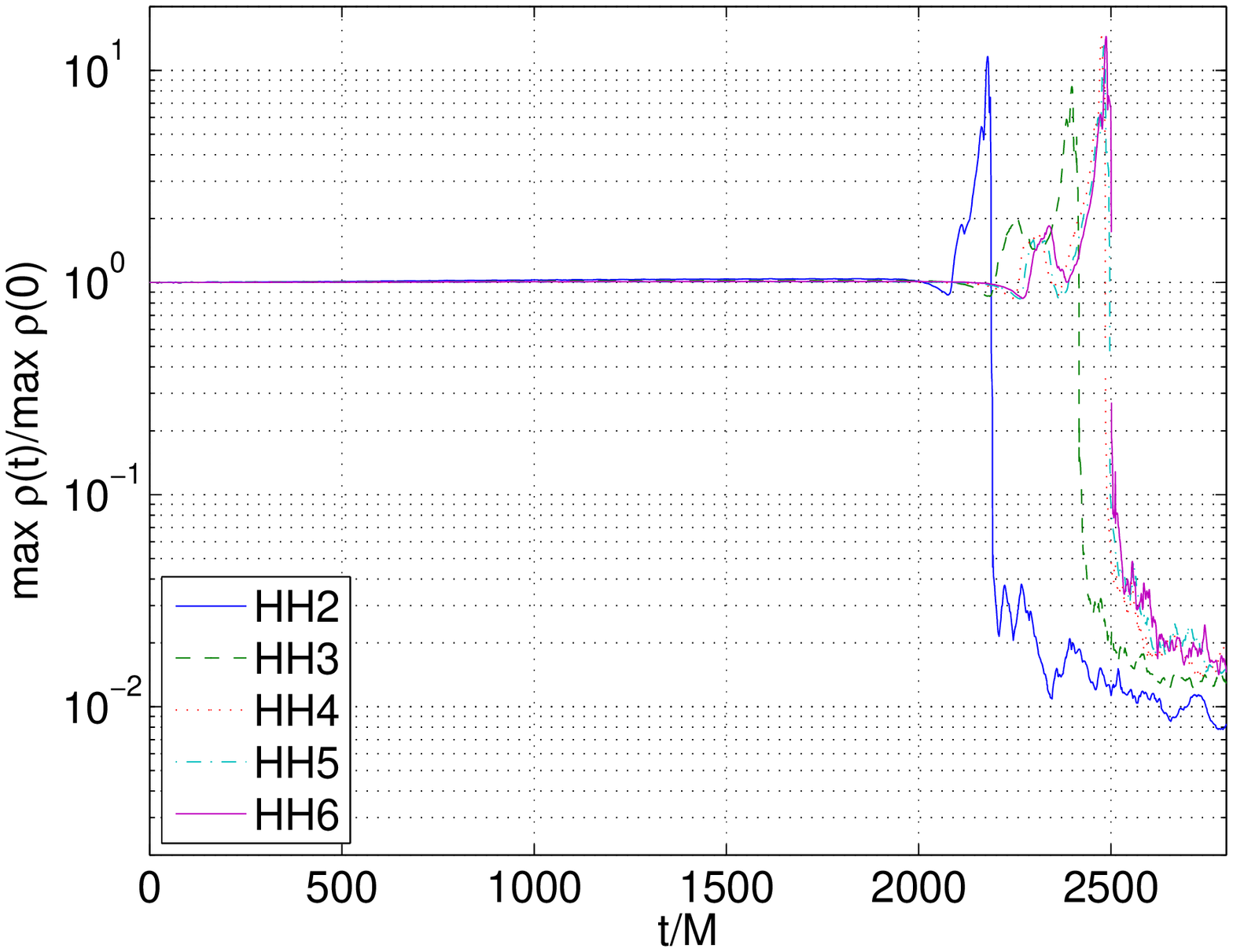}
    \includegraphics[width=0.49\textwidth]{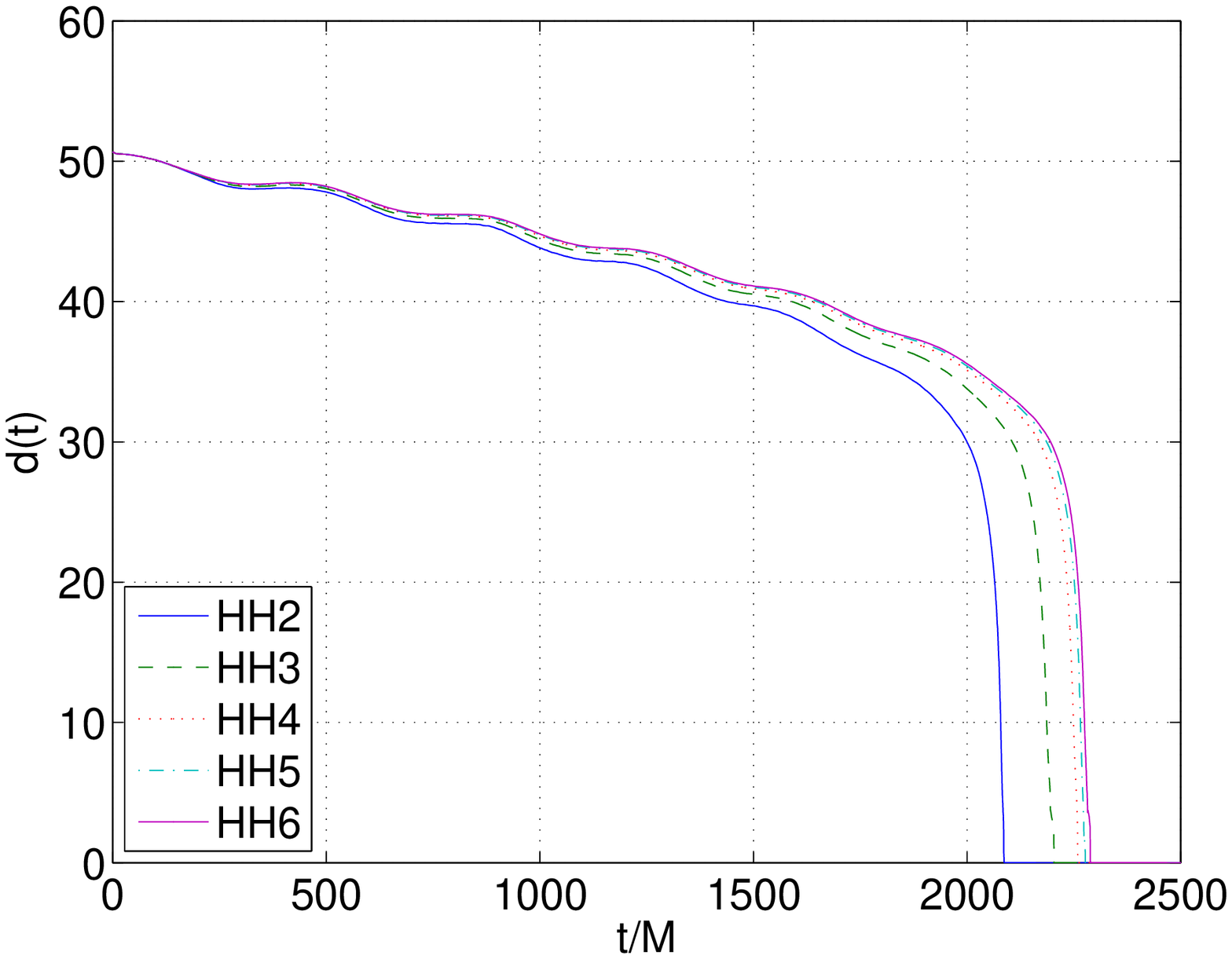}
    \caption{ \label{fig:dyn} BNS dynamics. 
      (left) Evolution of the maximum of the
      rest-mass density for different resolutions.
      (right) Evolution of the proper separation between the two stars.}
  \end{center}
\end{figure*}

The evolution of the maximum rest-mass density is reported in
Fig.~\ref{fig:dyn} (left), together with the proper distance (right). 
During the inspiral the maximum of the rest-mass remains constant as
expected; the proper distance shows some residual eccentricity from
the initial data. The latter is larger during the first three orbits
and then progressively radiated away, although not completely.
The stars touch each other about $1.5$ orbits before
merger ($t/M\gtrsim2050$), which happen at $t_{\rm m}/M=2259$ for run HH6 (see
Sec.~\ref{sec:waves} for the definition of merger used also in this work).
After the merger the maximum of the rest-mass density increases
indicating the compactness of the HMNS 
increases; it reaches a peak during the collapse then drops down 
to the densities of the accretion disk. The quasi-radial oscillations
of the HMNS are also visible before the collapse. 
An apparent horizon is formed at $t_{\rm AH}/M\simeq2475$ (run HH6),
the mass and spin of the final puncture describing the black
hole~\cite{Thierfelder:2010dv,Brown:2007tb,Hannam:2006vv} are $M_{\rm
  BH}=2.955\pm0.005$ and $a_{\rm BH}=0.80\pm0.01$. The
latter values are computed from the irreducible mass and spin, after
an initial transient in the BH formation.

Overall the new simulations at higher resolution confirm our previous
findings about the merger outcome~\cite{Thierfelder:2011yi}: the HMNS experiences a
delayed collapse~\footnote{ 
  In case thermal effect are included the HMNS survives even longer (several
  rotational periods) because of the additional pressure support due
  to temperature.} 
while a prompt collapse seems an artifact of lower resolution
runs (see HH2). Note also that the use of lower resolutions results in earlier
mergers. 

\begin{figure}[t]
  \begin{center}
    \includegraphics[width=0.49\textwidth]{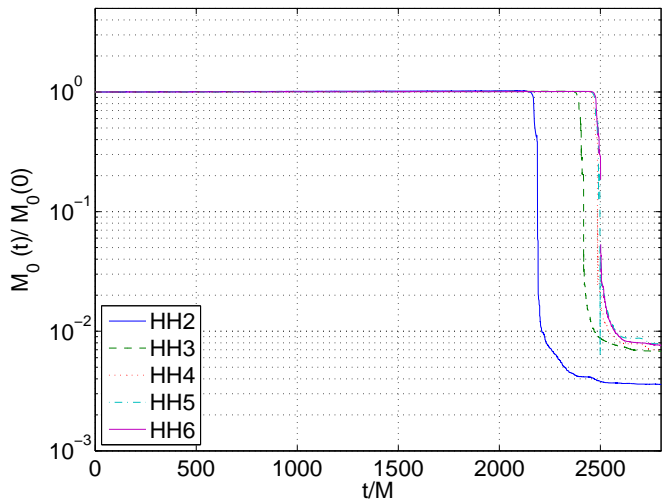}
    \caption{ \label{fig:m0} Evolution of the rest-mass density
      for different resolutions.} 
  \end{center}
\end{figure}

The evolution of the rest-mass is reported in Fig.~\ref{fig:m0}. 
During the inspiral it is conserved up to $\max(\Delta
M_0/M_0)\lesssim1\%$ for runs HH4 and higher resolutions. At the
collapse it drops several order of magnitude, similarly to the
maximum of rest-mass density. As explained
in~\cite{Thierfelder:2010dv} this effect is produced by the gauge 
conditions which, handling the singularity formation, stretch the
numerical grid effectively moving grid points to larger proper radii. 
The values of $M_0$ at late times are an estimate (upper limit) for
the rest-mass of the accretion disk, which is below $1\%$ of the
initial rest-mass. Note the strong dependence of the result on the
resolution.  

\begin{figure}[t]
  \begin{center}
    \includegraphics[width=0.49\textwidth]{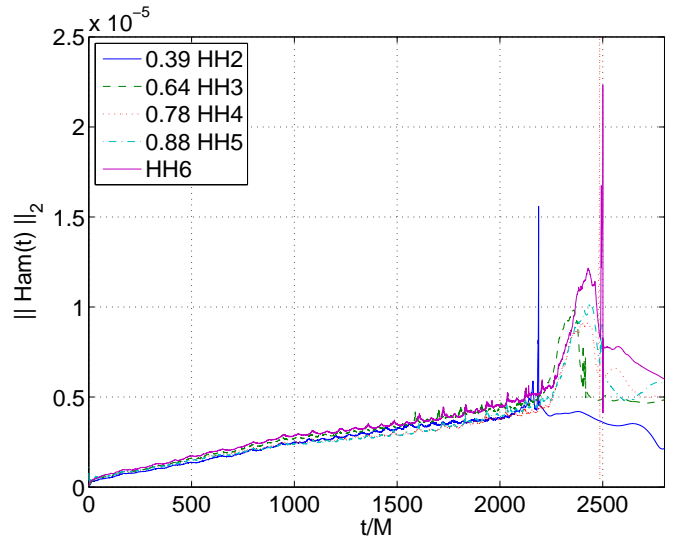}
    \caption{\label{fig:ham_norm} Convergence of the L2 norm of the
      Hamiltonian constraint. The data set are rescaled for second
      order convergence to the highest resolution one.}
  \end{center}
\end{figure}

The convergence of the L2 norm of the Hamiltonian constraint is
reported in Fig.~\ref{fig:ham_norm}. The different data set are
rescaled for second order convergence to the highest resolution
one. As one can observe from the figure, the lines are superposed
during the inspiral while they progressively differ from the contact,
towards the merger. After the merger convergence is not measurable.

Finally we comment about the ADM mass conservation.  We computed
finite-radius approximations to the ADM mass, $M_{\rm ADM}(r)$, by
integrals over coordinate spheres as in the case of the GW,
considering two different formulas: (a)~Eq.~(54)
of~\cite{Brugmann:2008zz}, and (b)~the integral of the conformal
factor only. The ADM mass is defined for the limit of large spheres,
$r\to\infty$. The value at finite $r$ is a coordinate dependent
quantity, and, for large but finite $r$, it suffers of resolution
problems in the outer levels.
In our setup the calculation is not accurate enough to make
quantitative statements and extrapolation $r\to\infty$ does not seem
to improve the results. We observe anyway a consistency between
$M_{\rm ADM}(r)$ and the energy of the emitted GW within the $1~\%$
level.

\section{Waveforms}
\label{sec:waves}

\begin{figure*}[t]
  \begin{center}
    \includegraphics[width=0.49\textwidth]{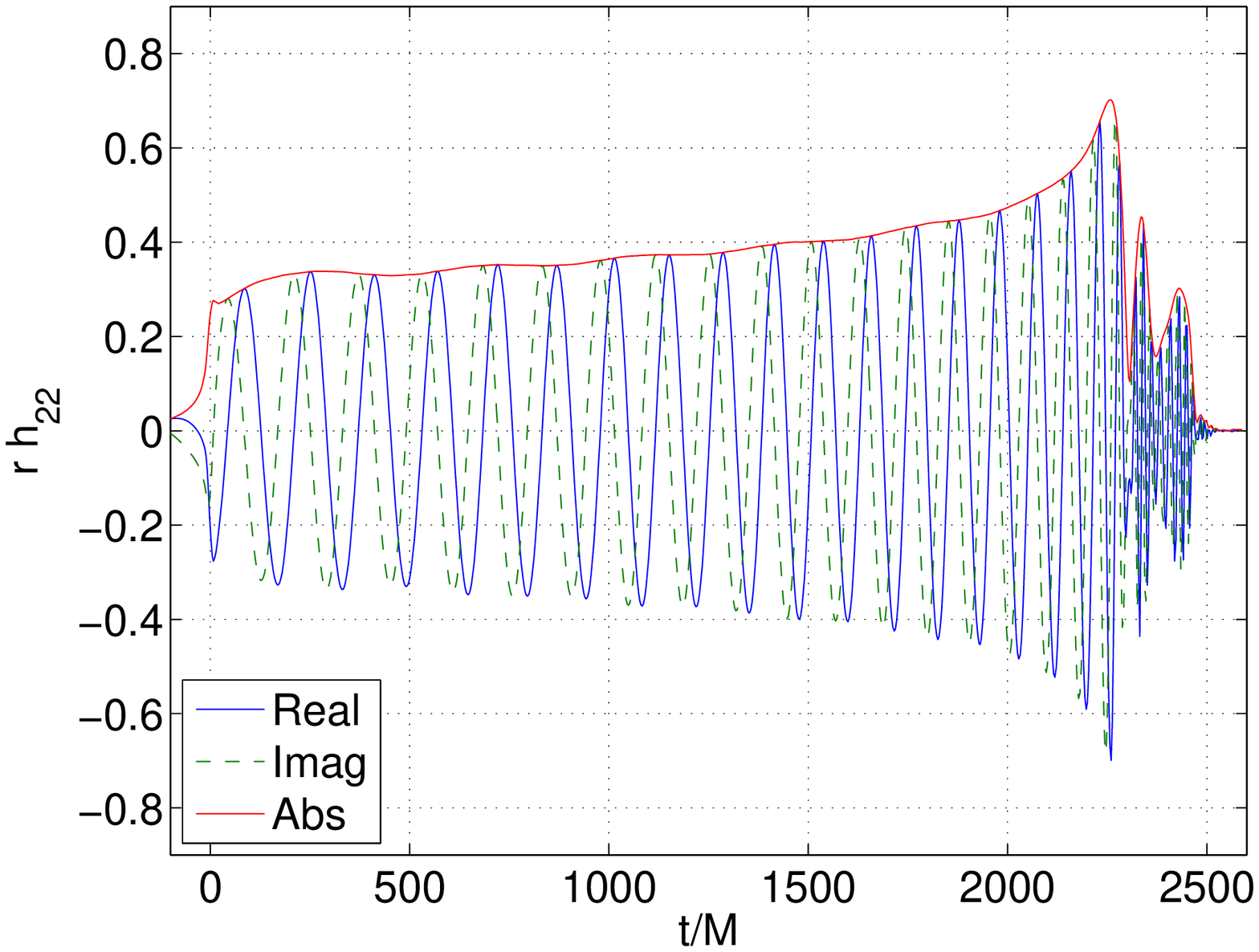}
    \includegraphics[width=0.49\textwidth]{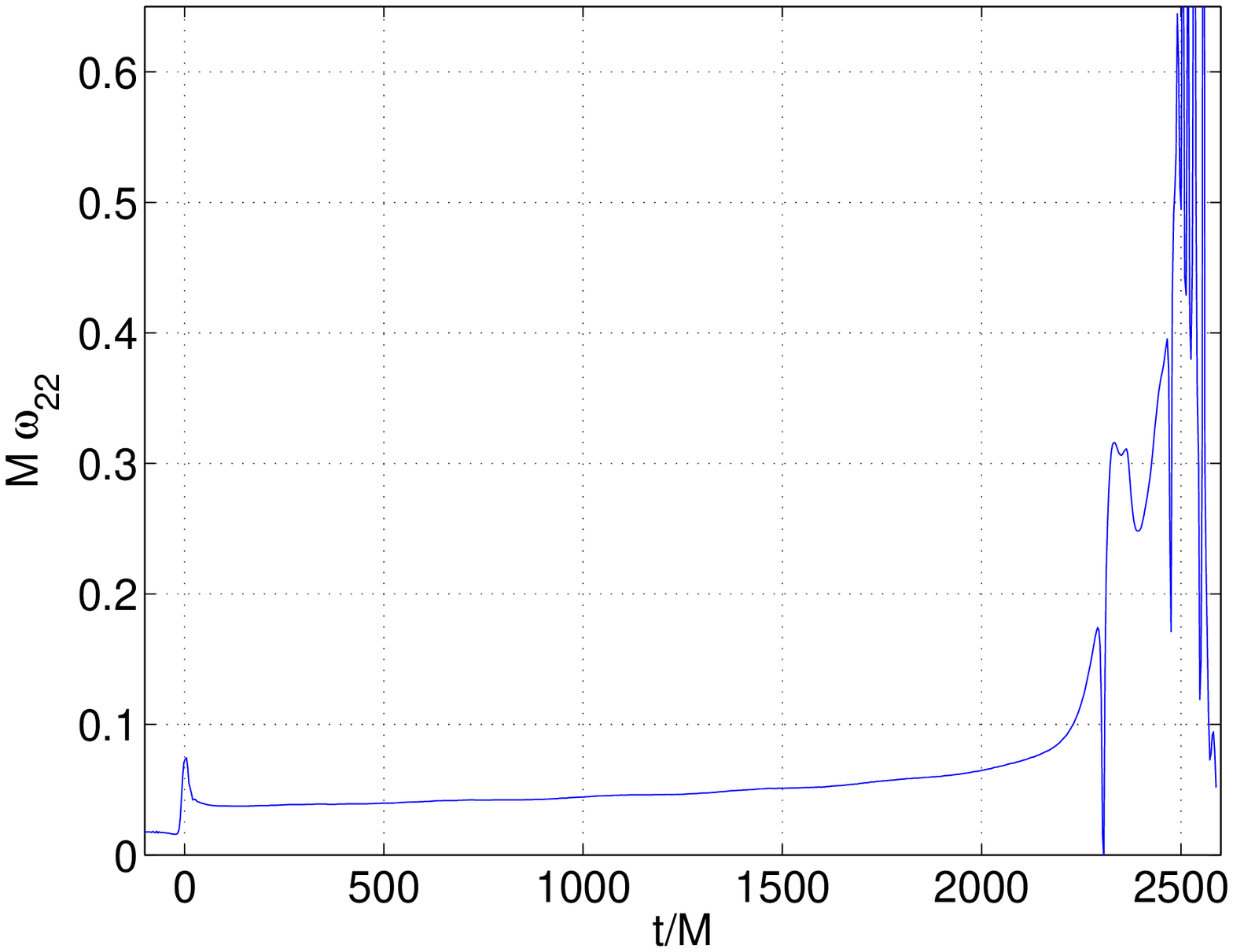}
    \caption{ \label{fig:h22} $r\,h_{22}$ waveform. 
      (left) Real part, imaginary part, and amplitude; (right)
      instantaneous GW frequency. Extraction radius $r=750$, run HH6.} 
  \end{center}
\end{figure*}

The total gravitational energy radiated during the merger process is
about $1~\%$ of the initial ADM mass. 
About $99~\%$ of the energy radiated during the inspiral is emitted into the
$(\ell,m)=(2,2)$ multipolar channel: the latter is also responsible
for about the $92~\%$ of the energy emitted during the whole
simulation. 
In the following we will consider only the $(2,2)$ mode. 
Figure~\ref{fig:h22} (left) shows the real part, the imaginary part, and the
absolute value of the GW multipole $r\,h_{22}$ extracted at $r=750$
($250~M$) and from the HH6 run. All the plots relative to the waveforms are in
term of the retarded time without changing notation. We use thus $t\to t-r_*$, 
where the tortoise radius is computed as $r_*=R + 2M\log(R/(2M)-1)$,
with $R(r)$ the Schwarzschild radius corresponding to the
coordinate extraction (isotropic) radius 
$r$. 

The waveform is characterized by the chirp-like shape typical of
the quasi-circular inspirals, after about $18$~cycles it peaks,
and then shows a more complicated structure with multiple maxima in
amplitude and progressively higher frequencies.
We formally define the merger time, $t_{\rm m}$, as the time
corresponding to the peak of the amplitude of  
$r\,h_{22}$~\cite{Baiotti:2011am,Thierfelder:2011yi}.
The signal after the merger is characterized by the emission from the
HMNS.  We reported an analysis in~\cite{Thierfelder:2011yi}, 
see also~\cite{Stergioulas:2011gd} for recent work. More details
will be given in a future work, in the following we will focus on
the inspiral waveforms.

The simulation output is the multipole of the curvature scalar,
$\psi^4_{22}$. The actual GW strain, $h$, is recovered from $\psi^4$ by
integrating the relation $\ddot{h}=\psi^4$. The integration is not
straightforward in the case of noisy numerical data. We employ the 
method described in~\cite{Reisswig:2010di}, i.e.~we perform the
integration in the Fourier frequency domain by applying a
fixed-frequency--high-pass filter, following
closely~\cite{Thierfelder:2011yi}.   
The cutting frequency used in this work is $f_{\rm cut}=0.0016<f_0$.
As shown in Fig.~\ref{fig:h22}, the waveform is affected by some
amplitude modulations mostly present at early times. Their origin
may be due to the residual eccentricity contained in the
initial data. 

In the following waveforms will be split into phase and amplitude 
according to the notation,  
\be
\label{eq:ap_notation}
r\,h_{22}=A_{22}\exp{(-i\Phi)} \ , \ \ \ 
r\,\psi^4_{22}= a_{22}\exp{(-i\phi)} \ .
\ee
The instantaneous GW frequency is $\omega=-\Im(\dot{h}/h)$, and is
plotted in Fig.~\ref{fig:h22} (right) in case of the $(2,2)$
multipole. It increases monotonically 
during the inspiral, and reaches the value $M\,\omega_{22}(t_{\rm
  m})\simeq0.123$ at the merger. At $t/M\sim2400$ it shows a signature of the
HMNS, and at later times increases to the quasi normal modes (QNMs)
frequencies of the final black hole (BH).
Note that the GW frequency drops to zero at $t/M\simeq2300$, corresponding to a
minimum of the amplitude and to a quasi spherical shape of the
stars~\cite{Thierfelder:2011yi}. The frequency of the BH fundamental
QNMs can be extracted from the GW frequency, however a cleaner
equivalent signal is provided by the frequency of $r\,\psi^4_{22}$.
We found for run HH6 $f_{\rm QNM}\simeq6.47~kHz$ ($M\,\omega_{22}\sim0.6$), 
in $2~\%$ agreement with the estimate obtained from the
horizon quantities.

In the following the accuracy of the numerical waveforms is assessed.
We stress that here, for the first time, phase and amplitude errors
are measured precisely and consistently from convergence tests.

\subsection{Convergence}
\label{sbsec:conv}

In this section we present the results concerning the self-convergence
of the inspiral waveforms. The convergence series HH$\{236\}$ is discussed 
as an example, similar results are obtained for HH$\{235\}$. We focus on
the extraction radius $r=750$ and on $r\,\psi^4_{22}$. Similar results
are found for $r\,h_{22}$.

\begin{figure*}[t]
  \begin{center}
    \includegraphics[width=0.49\textwidth]{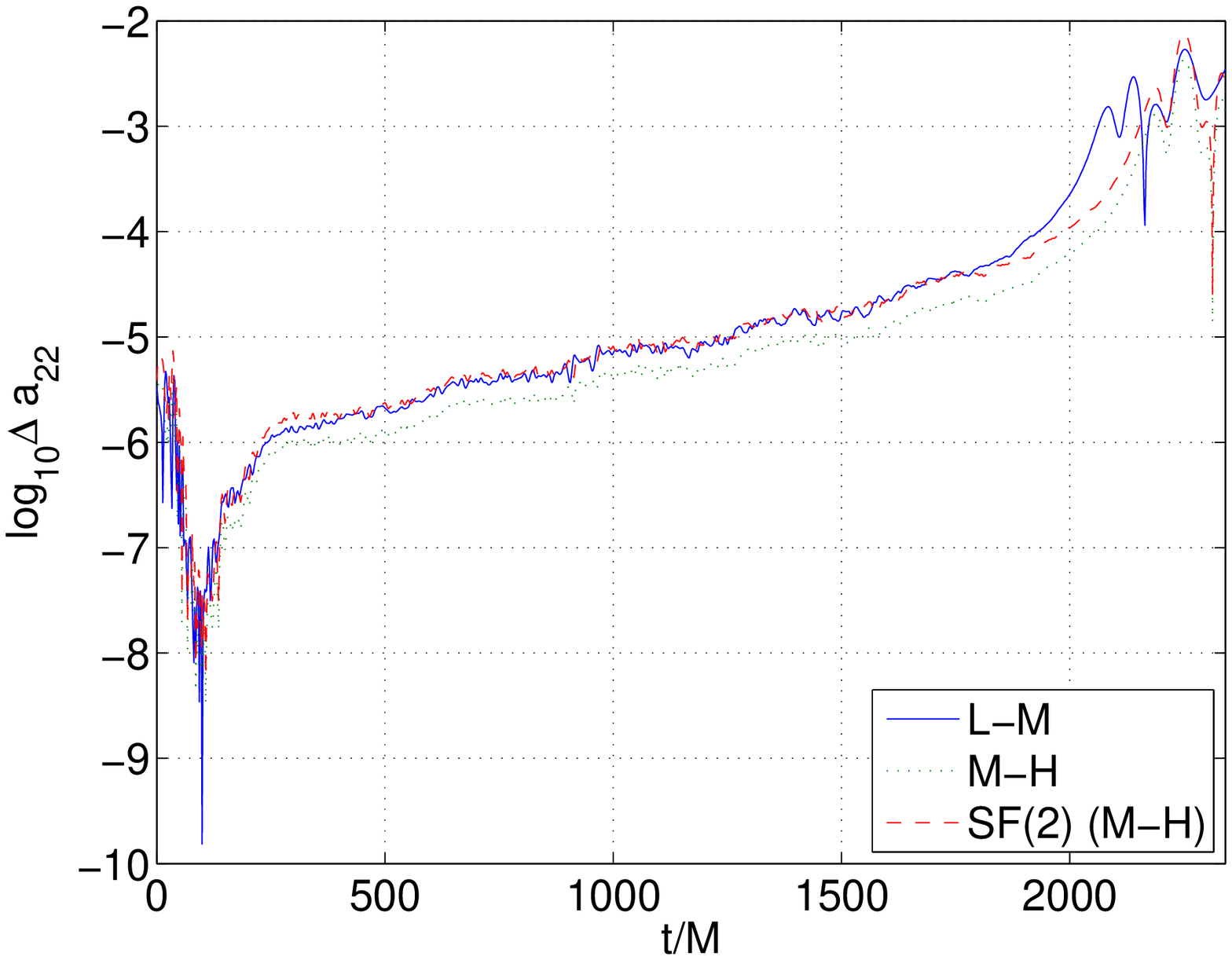}
    \includegraphics[width=0.49\textwidth]{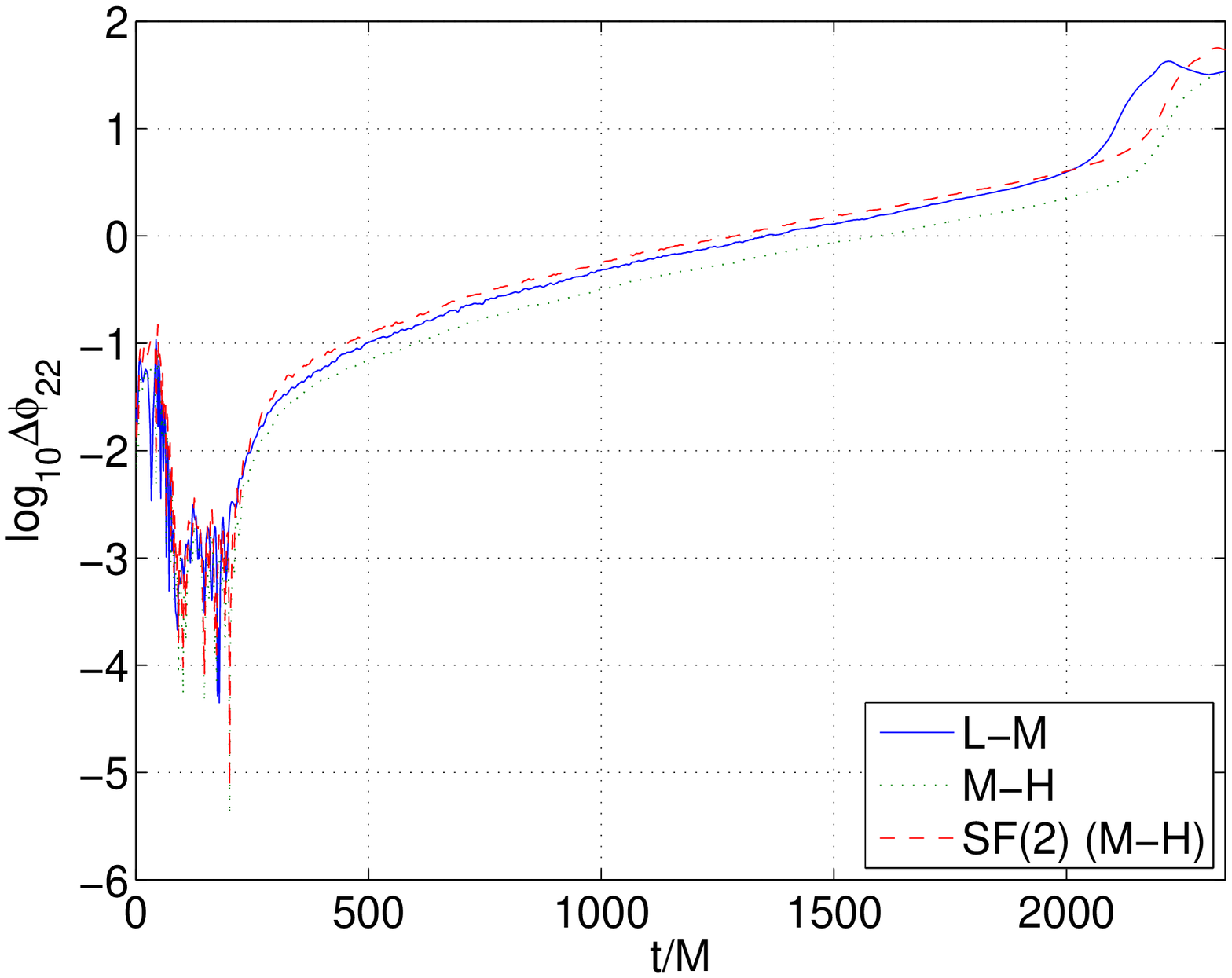}
    \caption{\label{fig:selfconv} Self-convergence of
      $r\,\psi^4_{22}$ waveforms. The logarithm differences between
      resolutions in amplitude (left) and phase (right) are shown
      together with the difference scaled for second order
      convergence. Extraction radius $r=750$.}   
  \end{center}
\end{figure*}

In Fig.~\ref{fig:selfconv} the self-convergence test is shown. 
The scaling factor is $SF(2)=1.8$. 
The differences are noisy so the figure employs a standard
Savitzky-Golay averaging filter for a better
visualization; results are not affected anyway.
The waveforms show compatibility with second order self-convergence during
the inspiral up to $t/M\simeq2000$. At later times they become, as
other quantities, progressively over-convergent, and after the merger
the convergence order can not be established. 
We observe here a common finding in NR simulations (e.g.~\cite{Yamamoto:2008js,Baiotti:2009gk,Thierfelder:2011yi}): 
as long as only the bulk motion of the matter is important, the
numerical methods employed do quite well modelling the inspiral due to
GW emission, but degrade when strong field and matter dynamics
develop.  

The over-convergence behavior appearing at late times is probably due
to the run HH2, but also to the fact that, when the stars come in
contact, the effective order of the (nonlinear) numerical scheme for
hydrodynamics probably drops below the second order (in norm).
As in previous shorter runs~\cite{Thierfelder:2011yi}, the
phase is not exactly convergent at rate two but at lower rate (between
one and two). 
Note that, differently from previous works on BNS and consistently
with~\cite{Thierfelder:2011yi},  
we do not align the waveforms for the convergence tests.
The gravitational energy carried by the $(2,2)$ mode
also shows approximately second order self-convergence. 

The interpretation of these data can be delicate because several
sources  of systematic errors are not completely under control: the 
exact expected convergence rate, the role of different grid setup, the
limited and not optimal choices of resolutions for the convergent
series,~etc. Our findings, however, appear consistent and sufficiently
robust; the second order rate is expected, in convergence regime, by
basic arguments, and the diagnostic quantities of Sec.~\ref{sec:dynamics}
show second order convergence in norm.
The results seem to indicate that second order convergence can be 
confidently assumed up to contact, or, equivalently, to
$M\,\omega_{22}=0.07$.  
In the following we will assume second order convergence for the
extrapolation of the inspiral waveforms up to merger, errors will be
given both for $M\,\omega_{22}\le0.07$ and for 
$M\,\omega_{22}\le0.1$. The reliability of the latter estimate is not clear.

\subsection{finite-radius extraction}
\label{sbsec:finiter}

In this section we study the uncertainties on phase and amplitude related to
the computation of waveforms at finite-extraction radii. 
We consider several extraction radii $r=200,\, 300,\, 400,\, 500,\,
750$ (or $R\simeq203,\, 303,\, 403,\, 503,\, 753$) from run HH6 and $r\,\psi^4_{22}$.

\begin{figure*}[t]
  \begin{center}
    \includegraphics[width=0.49\textwidth]{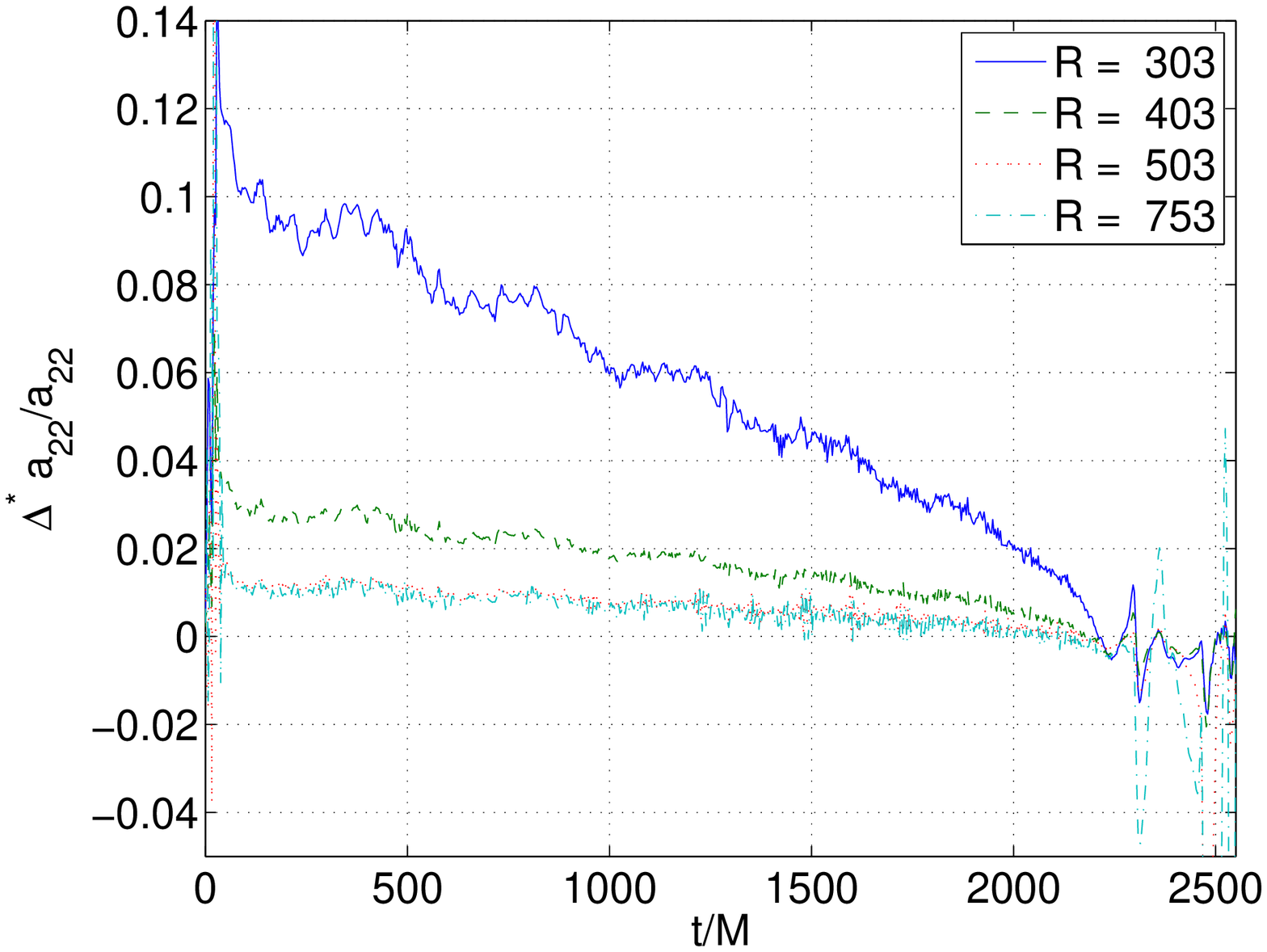}
    \includegraphics[width=0.49\textwidth]{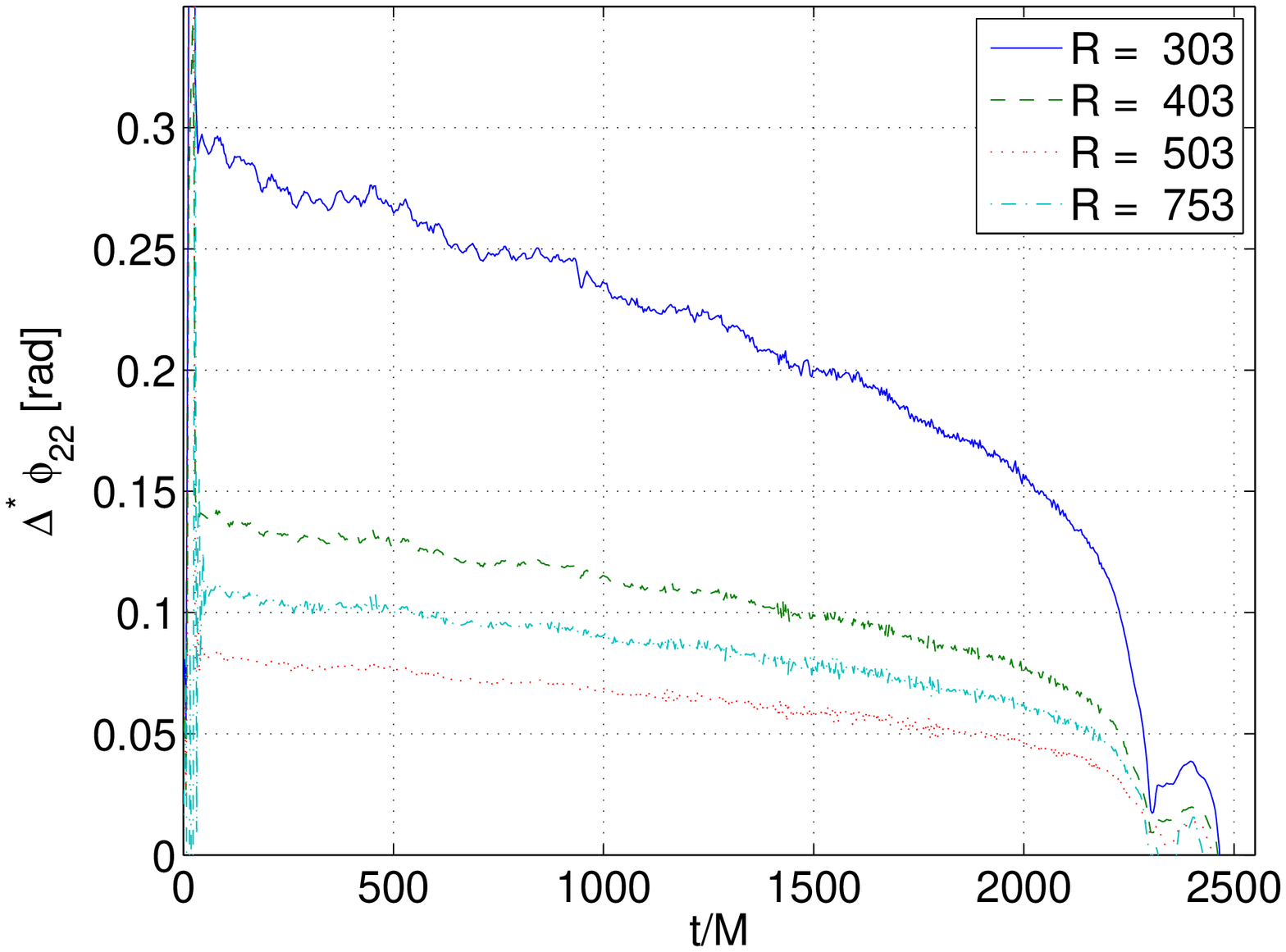}
    \caption{\label{fig:finiter:diffs} Differences between 
      amplitude (left) and phase (right) of waves 
      extracted at successive radii $r=200,\, 300,\, 400,\, 500,\, 750$.
      Run HH6.}
  \end{center}
\end{figure*}

The differences in amplitude and phase extracted at a given radius
with the previous,
e.g.~$\Delta^*\phi_{22}(R_i)=\phi_{22}(R_i)-\phi_{22}(R_{i-1})$,  are
shown in Fig.~\ref{fig:finiter:diffs}.
Both amplitude and phase increase for higher extraction radii.
The differences are bigger at earlier times, the phase differences at
early times scale approximately as $1/r$, while amplitude differences
approximately as $1/r^2$. 
For radii $r\geq400$ differences in amplitude are $\lesssim2\%$ and
they seem to saturate. By contrast differences in phase keep on
increasing  and between $r=750$ and $r=500$ they are
$\sim0.1$~$\rad$. The differences become progressively smaller towards
the merger.    

\begin{figure}[t]
  \begin{center}
    \includegraphics[width=0.49\textwidth]{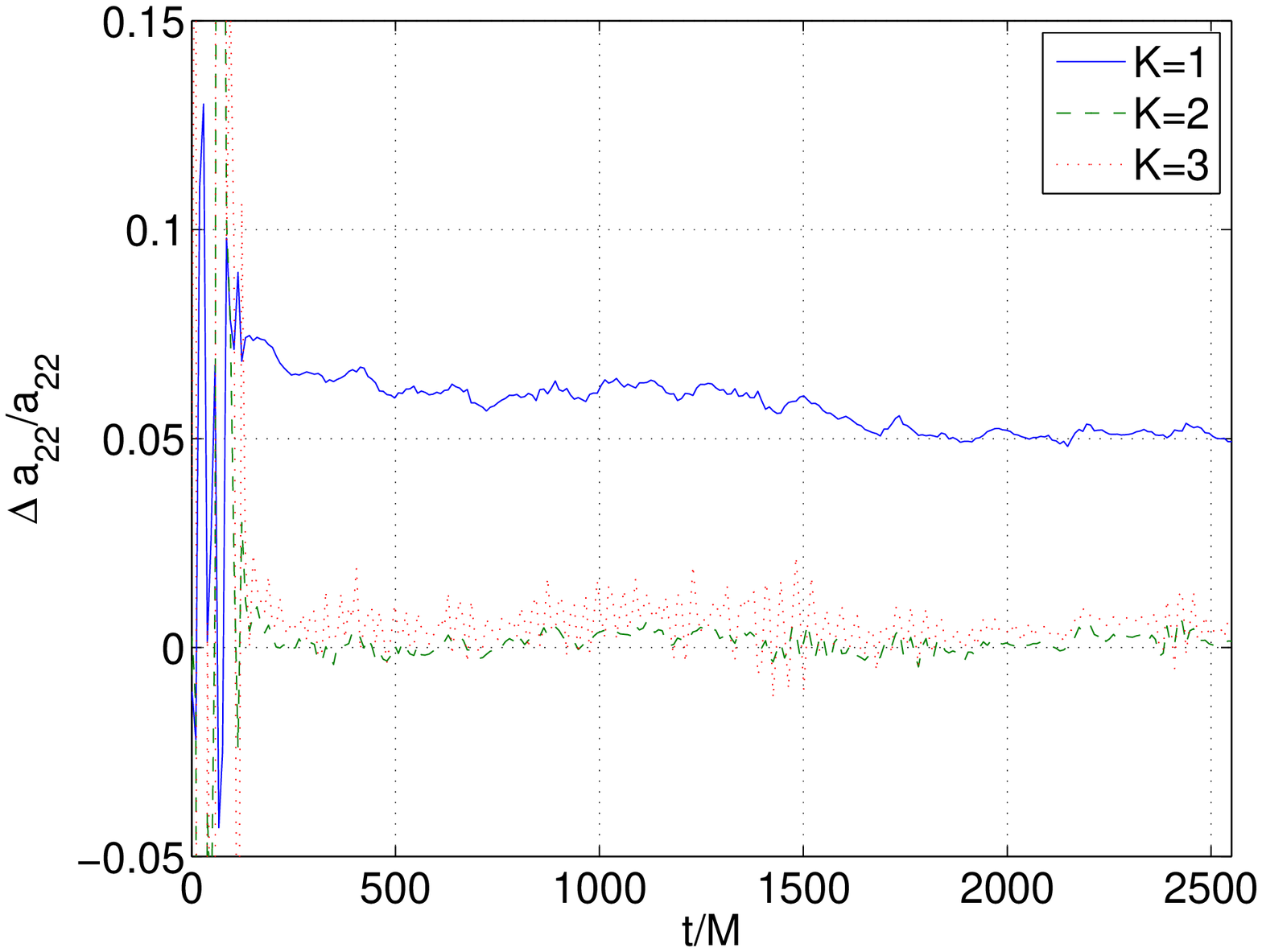}
    \includegraphics[width=0.49\textwidth]{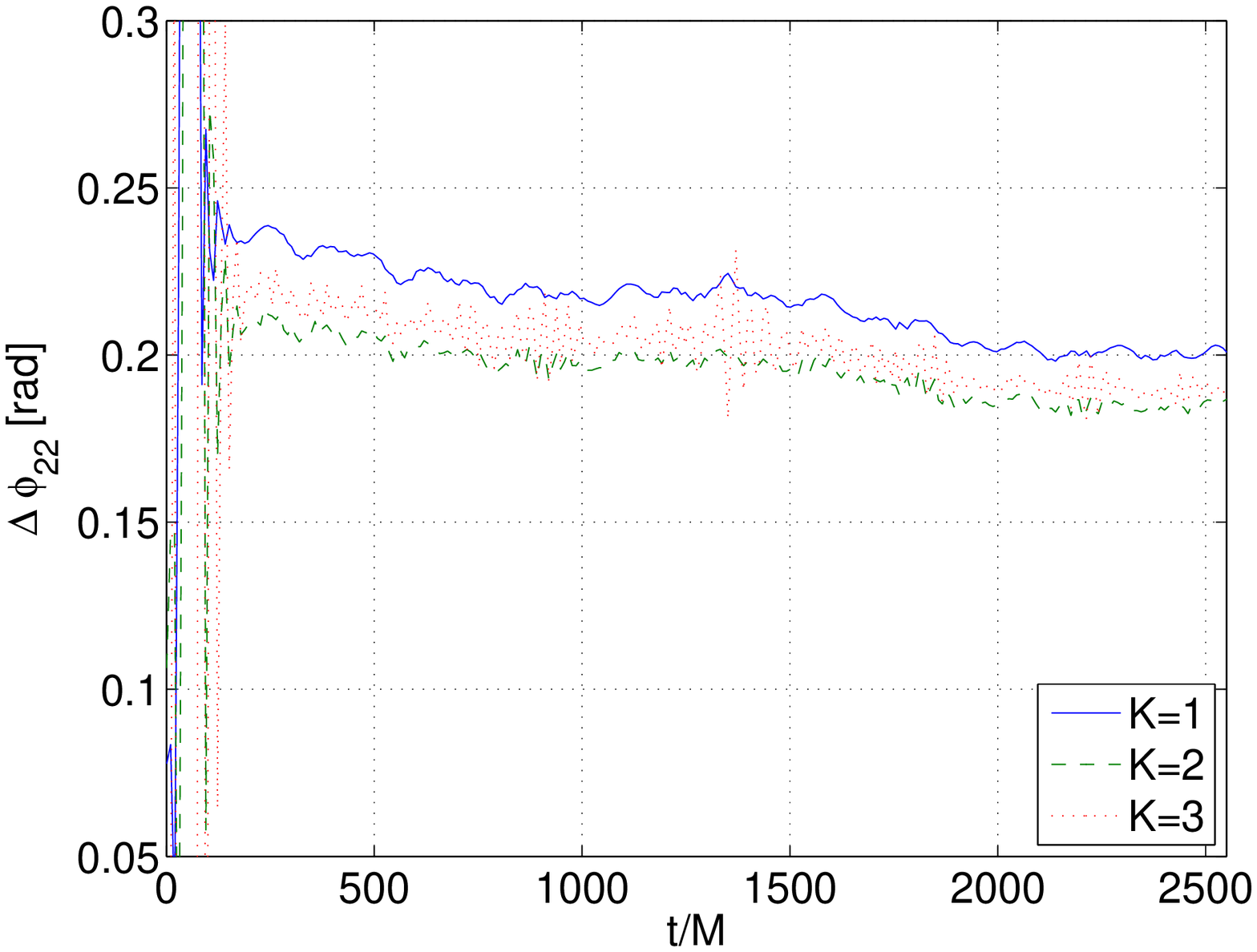}
    \caption{\label{fig:finiter:extrap} Differences between amplitude
      (left) and phase (right) of $r\,\psi^4_{22}$ extrapolated up to 
      different orders, $K$, and at the reference radius $r=750$. Run HH6.} 
  \end{center}
\end{figure}

Following previous
works~\cite{Boyle:2007ft,Scheel:2008rj,Boyle:2008ge,Boyle:2009vi,Pollney:2009ut,Pollney:2009yz},
an approximation of the waves at null-infinity can be obtained by simple $1/R$-extrapolation, 
\be
F(t,R) = \sum_{k=0}^{K} F_k(t) R^{-k} \ ,
\ee
where $F(t,R)$ is either the phase or the amplitude, $t$ is the
retarded time, and $F_0(t)$ is the extrapolated value. 
In~\cite{Reisswig:2009us,Reisswig:2009rx,Babiuc:2011qi}
the robustness of the extrapolation procedure has been assessed
against null-infinity waveforms from equal-masses BBH inspirals
computed with the Cauchy-characteristic extraction (CCE)
method~\cite{Bishop:1996gt,Babiuc:2005pg,Babiuc:2008qy}. 
In~\cite{Bernuzzi:2010xj,Bernuzzi:2011aj}, by mean of BH perturbation
theory on hyperboloidal foliations, it has been shown that, in case of an
unambiguous definition for the background and for the retarded time,
the extrapolation reproduce null-infinity waveforms up to their
numerical uncertainties for enough high values of $K>3$.

Figure~\ref{fig:finiter:extrap} shows the differences between the
extrapolated value for different $K$ and the reference radius 
$r=750$. 
Note that waveforms at different radii are not shifted in time or
phase, but only considered against the retarded time.
The fit errors, $\delta F$, are computed at the $68~\%$ confidence
level, and distributed quite uniformly in the inspiral. Hence, 
their average, $\avg{\delta F}$, can be use as a meaningful measure
of the fit quality in comparison with the differences, $\Delta F$,
between the extrapolated waves and the finite-radius extracted ones.  
In case of a linear extrapolation, $K=1$, the average fit errors are 
$\avg{\delta\phi_{22}}\sim0.02$~$\rad$ and
$\avg{\delta a_{22}/a_{22}}\sim3~\%$, and the maximum
differences with the last radius are $\Delta a_{22}/ a_{22}\sim+5~\%$ and
$\max\Delta\phi_{22}\sim+0.23$~$\rad$.  
For $K=2$, the average fit errors are 
$\avg{\delta\phi_{22}}\sim0.04$~$\rad$ and
$\avg{\delta a_{22}/a_{22}}\sim3~\%$, and maximum
differences with the last radius are $\max\Delta a_{22}/a_{22}\sim+1\%$ and
$\max\Delta\phi_{22}\sim+0.21$~$\rad$. The use of $K>2$ results in more
noisy data as shown by the figure, and also the fit errors increase.
The ``best'' extrapolation is thus given by $K=1$ or $K=2$. Note
however that the fit averaged uncertainties are about $10~\%$ of the
phase difference with the last resolution also in the best cases, and
that, within this uncertainty, both 
the extrapolations basically agree (see right-hand panel of
Fig.~\ref{fig:finiter:extrap}). 

\begin{figure}[t]
  \begin{center}
    \includegraphics[width=0.49\textwidth]{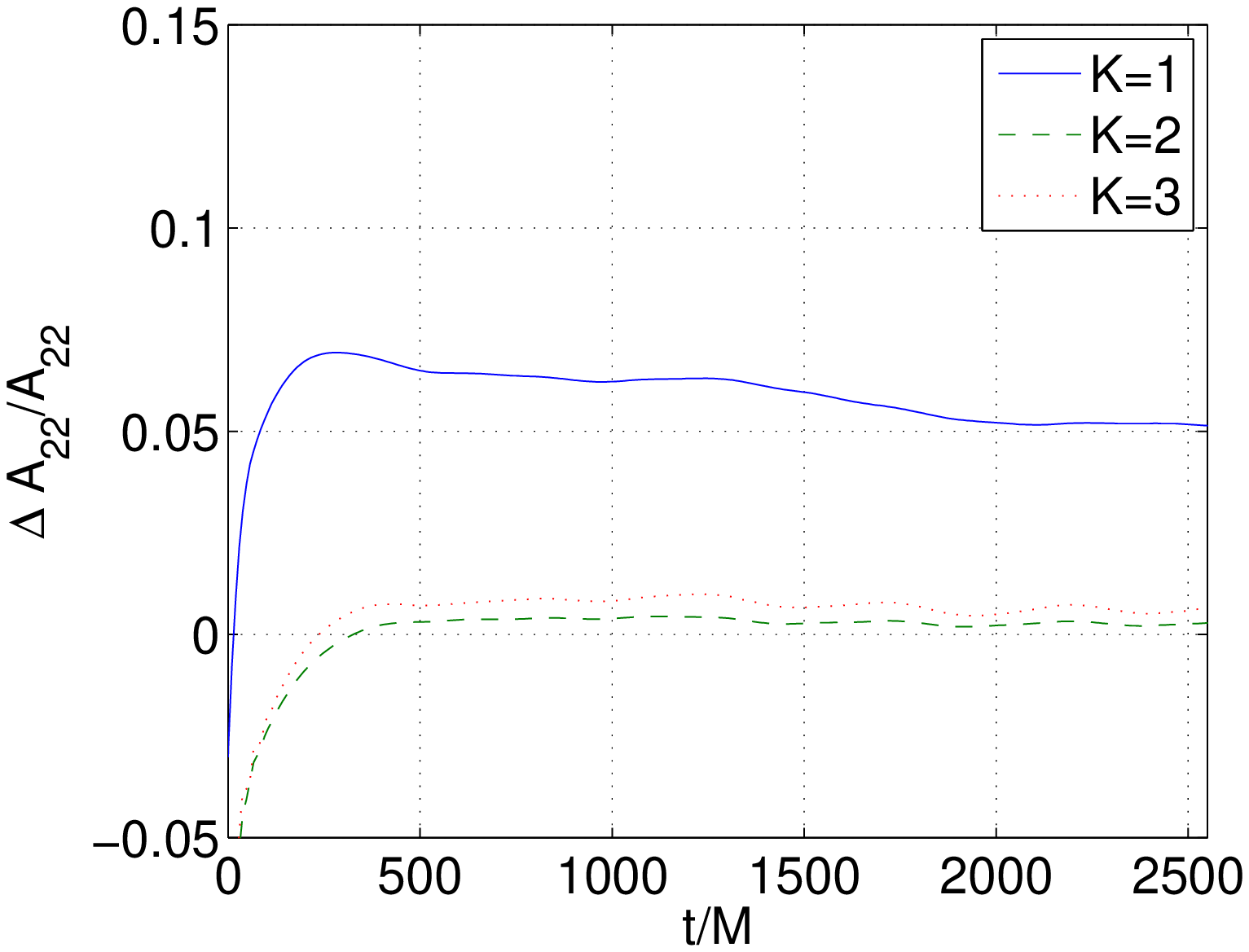}
    \includegraphics[width=0.49\textwidth]{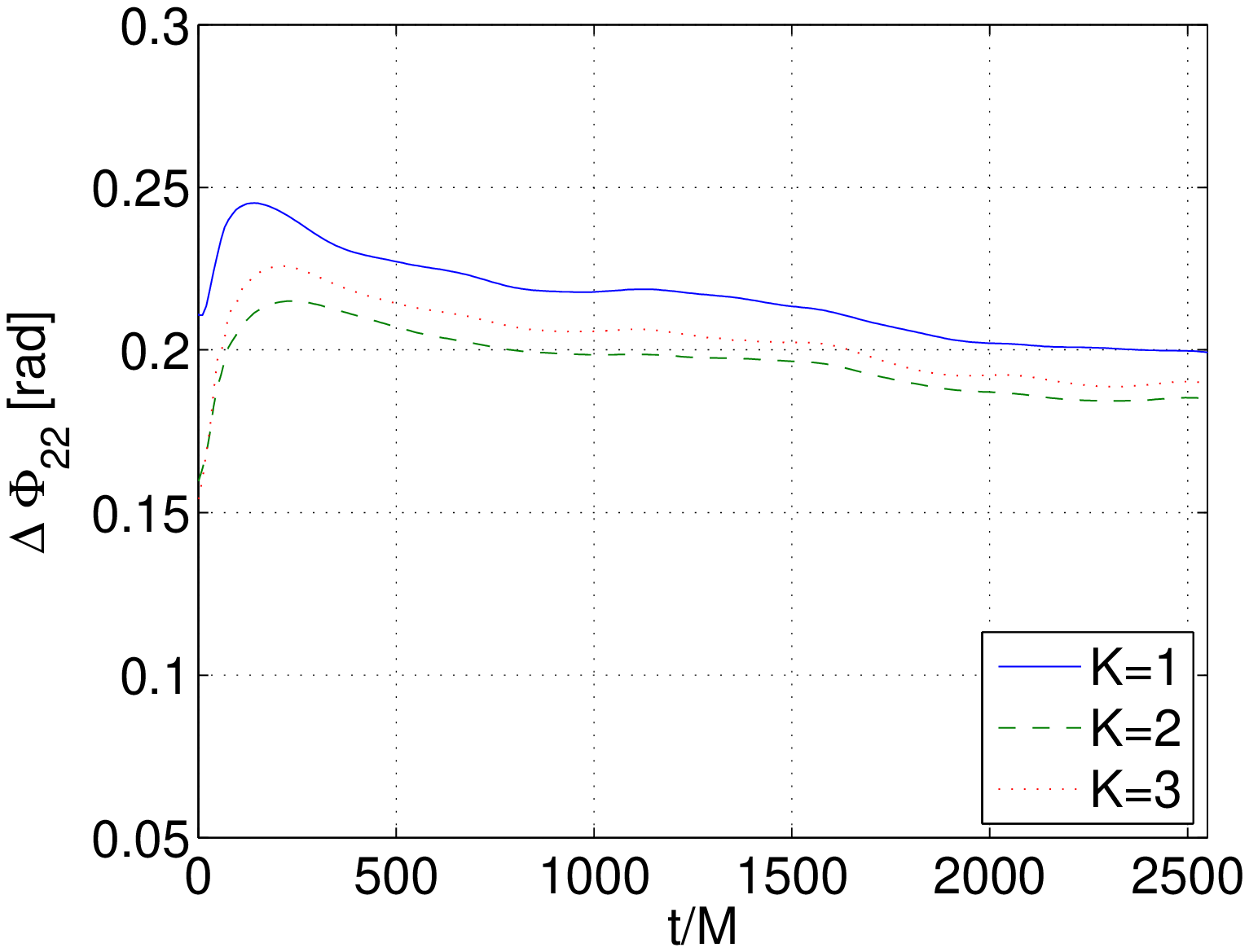}
    \caption{\label{fig:finiter:extraph} As
      Fig.~\ref{fig:finiter:extrap} but referring to $r\,h_{22}$.}
 \end{center}
\end{figure}

A similar behavior has been observed for the
extrapolation of $r\,h_{22}$. In this case however data are less noisy
and the fit errors are smaller. Specifically we found 
$\avg{\delta\Phi_{22}}\sim0.006$~$\rad$ and
$\avg{\delta A_{22}/A_{22}}\lesssim2~\%$ for $K=1$ and 
$\avg{\delta\Phi}_{22}\sim0.002$~$\rad$ and
$\avg{\delta A_{22}/A_{22}}\lesssim1~\%$ for $K=2$. The latter
is thus preferable. The differences with the last extraction radius
are reported in Fig.~\ref{fig:finiter:extraph}, and comparable in
absolute size to those of Fig.~\ref{fig:finiter:extrap}.

\subsection{Truncation errors}
\label{sbsec:trunc}

\begin{figure}[t]
  \begin{center}
    \includegraphics[width=0.49\textwidth]{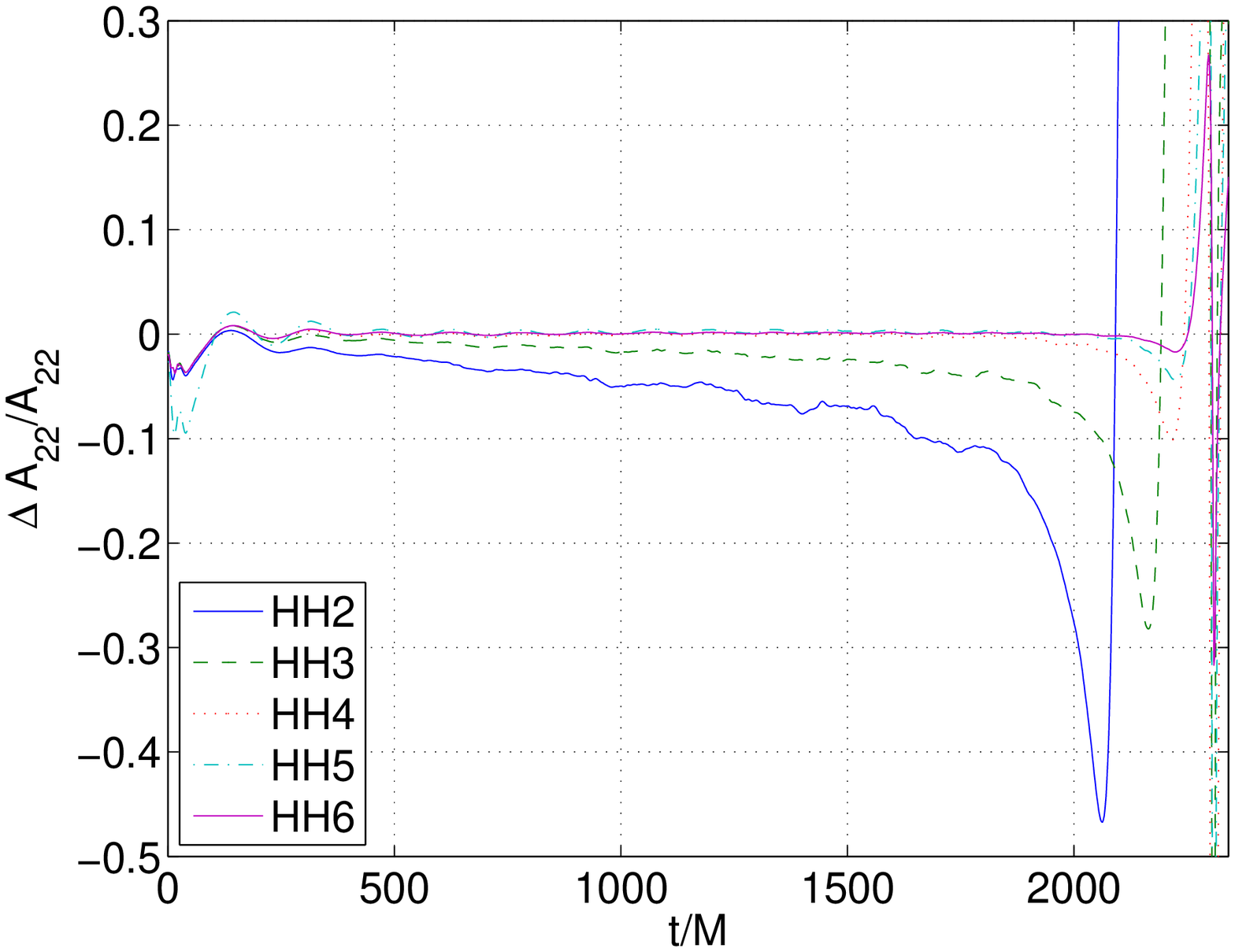}
    \includegraphics[width=0.49\textwidth]{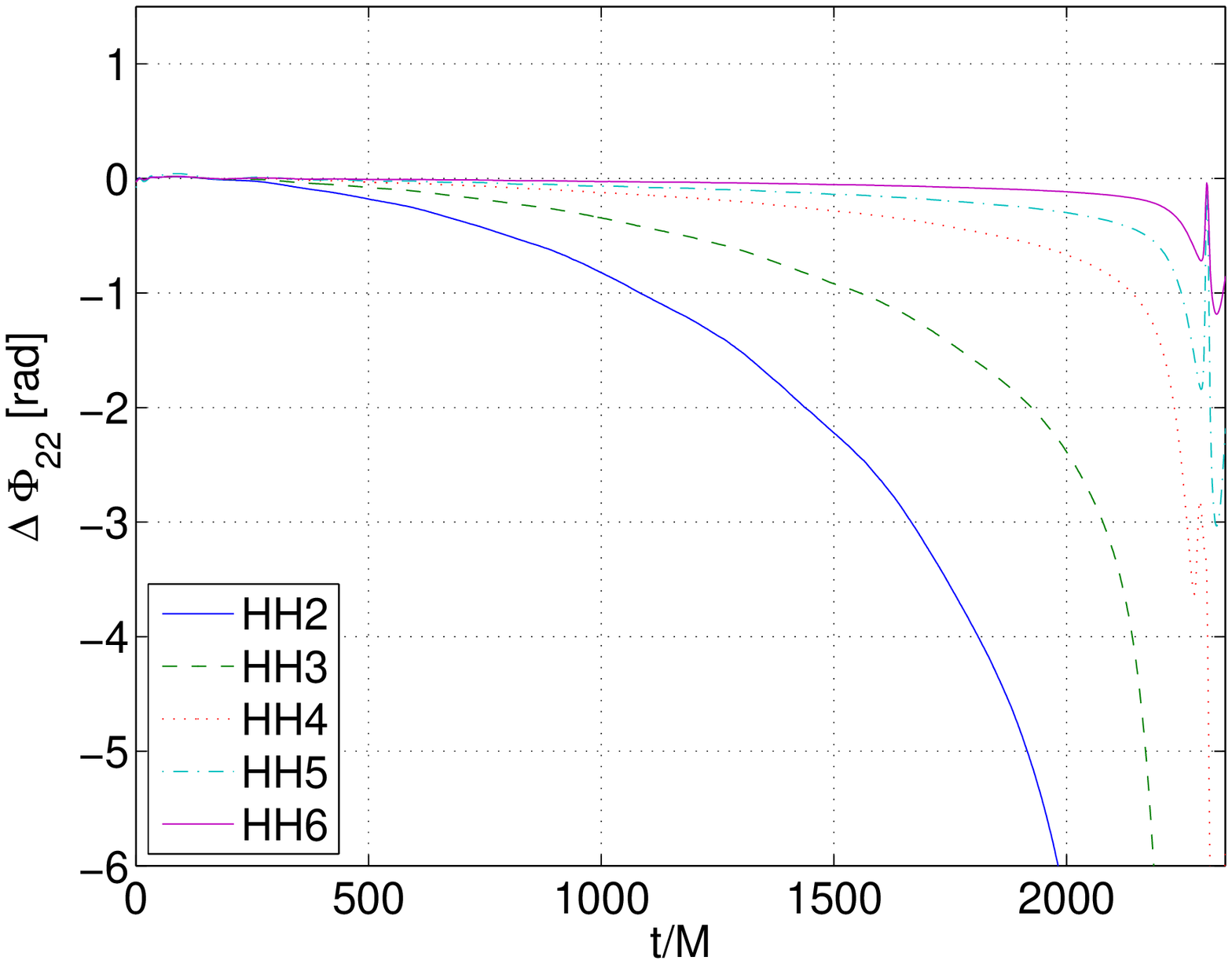}
    \caption{\label{fig:diffextrp} Differences between extrapolated
      values from series HH$\{23456\}$ and the data at various
      resolution. Amplitude (left) and phase (right) of $r\,h_{22}$
      (bottom). Extraction radius $r=750$.} 
  \end{center}
\end{figure}

In this section we quantify the truncation errors in the inspiral waveforms.
Richardson extrapolation is employed using different data sets and assuming second
order convergence. Extrapolation series are indicated with the same
notation of convergence series, e.g.~HH$\{23456\}$. Errors are computed as
differences with the highest resolution data.
We stress that this is a common but optimistic choice.
Also, to avoid underestimates of the errors, the whole convergent
series is, at least, used in the extrapolation. 

\begin{table*}[t]
  \caption{\label{tab:maxerrs} Maximum differences between extrapolated
    values in resolution and the highest resolution data (run HH6) in
    different quantities during the inspiral. The first three rows
    refer to the maximum errors for $M\,\omega_{22}\le0.07$, while the
    last three rows refer to the maximum errors for
    $M\,\omega_{22}\le0.1$.} 
  \centering    
  \begin{tabular}{ccccccc}        
    \hline
    Runs 
    & $|\Delta a_{22}/a_{22}|$ [\%] & $|\Delta \phi_{22}|$ [$\rad$] 
    & $|\Delta A_{22}/A_{22}|$ [\%] & $|\Delta \Phi_{22}|$ [$\rad$] 
    & $|\Delta \omega_{22}/\omega_{22}|$ [\%]\\
    \hline
    HH$\{236\}$   & 38 & 1.8  & 8   & 2    & 9 \\
    HH$\{2346\}$  & 5  & 0.4  & 1   & 0.4  & 2 \\
    HH$\{23456\}$ & 1  & 0.13 & 0.2 & 0.13 & 0.6 \\
    \hline                                   
    HH$\{236\}$   & $>$100 & 7   & 20 & 5   & $>$100 \\
    HH$\{2346\}$  & 70     & 1.4 & 6  & 1.4 & 15 \\
    HH$\{23456\}$ & 13     & 0.3 & 2  & 0.3 & 4 \\
    \hline

  \end{tabular}
\end{table*}

Figure~\ref{fig:diffextrp} shows the differences in amplitude and
phase between the extrapolated data from HH$\{23456\}$ and those from run HH6. 
Similar plots were produced for
$r\,\psi^4_{22}$ and for different extrapolation series. 
The differences are negative and increase towards
the merger. Before the merger the trend changes and they rapidly
increase to positive values. From the argument given at the end of
Sec.~\ref{sbsec:conv}, we do not expect to have a fully reliable
extrapolation at the merger, thus proper error estimates must be
restricted to slightly before that point. 
In case of other extrapolation series the errors
are bigger but qualitatively they show the same behavior.
The maximum absolute errors observed before the merger 
are reported in Tab.~\ref{tab:maxerrs} for different extrapolation series.
They are computed within the GW frequency intervals $M\,\omega_{22}=[0.0358,0.07]$
(i.e.~$f\in[f_0,f_{\rm max}]=[0.0019,0.0037]$), where the
extrapolation is reliable, and also within the GW frequency intervals 
$M\,\omega_{22}=[0.0358,0.1]$. In the latter case they roughly
correspond to the minima in Fig.~\ref{fig:diffextrp}.

As shown by the table, it is necessary to include at least four 
resolutions to obtain a phase error $\Delta\Phi_{22}\lesssim1$~$\rad$ and
an amplitude error of $\Delta A_{22}/A_{22}\lesssim1~\%$ for
$M\,\omega_{22}\le0.07$. In this case truncation errors become of the
same order of magnitude of the finite-extraction effect. 
The error estimate up to $M\,\omega_{22}=0.1$ indicate how
dramatically the errors increase up to merger.
This analysis suggests that truncation errors represent the main
source of uncertainties in BNS simulations.

\subsection{Accuracy}
\label{sbsec:acc}

In this section we test the inspiral waveforms against accuracy
standards for data analysis. 
As a measure of the accuracy we employ the square
of the inaccuracy functional, $\I^2$, whose definition is discussed in
detail in Appendix~\ref{app:astd}, Eq.~\eqref{eq:I2}. Accuracy
requirements are set to minimal levels and an ideal detectors is
assumed. 
As mentioned in Appendix~\ref{app:astd}, two waveforms are
distinguishable depending on the signal-to-noise ratio (SNR, $\vrho$) of the
detection. Given a difference between two waveforms (the error bars,
in our case), $\delta h$, there always exists a sufficiently high SNR 
such that the difference is significant (in our case, the waveforms
are inaccurate). The point is thus to assess the accuracy with respect
SNR that are high enough but also realistic for the future detections.
We recall that for equal-masses BBH waveforms accuracy
standards are achieved for relevant SNR, and waveforms can be
considered faithful,
e.g.~\cite{Aylott:2009ya,Hannam:2009hh,Hannam:2010ky,MacDonald:2011ne}.  
By contrast, such analysis for BNS has never been considered before. 

We recall that the inaccuracy functional has been rarely 
employed in NR literature, see e.g.~\cite{MacDonald:2011ne}, 
but it provides an equivalent measure to the most common mismatch functional,
$\M$ (see again Appendix~\ref{app:astd} for the definition).
The inaccuracy functional is here preferred because its value does not
depend on the distance between the detector and the source or
on the normalization of the PSD of the detector noise.
In the accuracy standards the dependency on the distance from the
source is then moved to the right-hand-side of the expressions.
All the results presented here can be translated in term of the latter
considering that $\I^2/\vrho\approx
2\M$~\cite{McWilliams:2010eq,Hannam:2010ky}. 

\begin{table*}[t]
  \caption{\label{tab:acc}   
    Inaccuracy functional for several configurations.
    Columns: detector configuration with 
    reference for the noise curve, 
    SNR at $100~Mpc$ (assuming the detector and the binary are
    optimally aligned),  
    inaccuracy functional for different choice of the waveforms.
    HH$\{2-6\}$ indicates the waveform has been extrapolated in resolution
    with those runs, ${}^*$ indicates the extrapolation in radius is
    performed. 
    The model waveform, $h_{\rm m}$, in the inaccuracy functional is
    always the one from run HH6 at $r=750$, except for the last column
    where the extrapolated in radius from the same run is used.}
  \centering    
  \begin{tabular}{l|c|c|c|c|c|c}        
    \hline
    $S_n$ & $\vrho$ & \multicolumn{4}{c}{$\I^2[h_{\rm m},h_{\rm x}]$}\\   
    & 
    & HH6, HH$\{236\}$
    & HH6, HH$\{2346\}$
    & HH6, HH$\{23456\}$
    & HH6, HH$\{23456\}^*$
    & HH$6^*$, HH$\{23456\}^*$ \\ 
    \hline    
    advLIGO~\cite{Ajith:2009fz} & 
    3.6  & 0.333 & 0.117 & 0.043 & 0.149 & 0.045\\
    
    advLIGO NSNS Opt~\cite{Sn:advLIGO} & 
    5.4  & 0.352 & 0.126 & 0.046 & 0.144 & 0.048\\
    advLIGO Narrow Band~\cite{Sn:advLIGO} & 
    3.2 & 0.462 & 0.198 & 0.072 & 0.112 & 0.074\\
    
    advLIGO High Sens.~\cite{Sn:advLIGO} & 
    5.1 & 0.401 & 0.153 & 0.055 &  0.132 & 0.058\\
    advVIRGO~\cite{Ajith:2009fz} & 
    5.1 & 0.445 & 0.182 & 0.066 & 0.117 & 0.068\\
    
    ET~\cite{Hild:2008ng,Sn:ETB} & 
    52.0 & 0.383 & 0.142 & 0.051 & 0.138 & 0.054\\
    \hline
  \end{tabular}
\end{table*}

In Tab.~\ref{tab:acc} we report, for several detector configurations,
the SNR, computed assuming the source at an effective
distance of $100~\Mpc$, and the inaccuracy functional, computed for
different waveforms and choices of the error bars.
Specifically, $\I^2[h_{\rm m},h_{\rm x}]$ is computed employing as
exact waveform ($h_{\rm x}$) three waveform extrapolated in
resolution, HH$\{236\}$, HH$\{2346\}$, and HH$\{23456\}$, and one
extrapolated in resolution \emph{and} radius using the series
HH$\{23456\}$ and the $K=2$ series of Sec.~\ref{sbsec:finiter}. 
The choice of the model waveform, $h_{\rm m}$, in $\I^2$ basically
determine the size of the uncertainties.
In the table we employ the waveform from the highest resolution run
HH6 and extracted at $r=750$, except for the last column which
employed the extrapolated in radius form the same run.  
Other choices were considered but they are not shown in the table.
The Wiener scalar product is computed on the interval
$f\in[f_0,f_{\rm max}]$ as in Sec.~\ref{sbsec:trunc}. Extrapolation in
radius is marked with ${}^*$.

As already clear from the analysis of Sec.~\ref{sbsec:trunc} the
inaccuracy decreases by increasing the number of runs used in the
extrapolation in resolution. The inaccuracy further increases if
finite-extraction effect are included. Let us consider the criteria in
Eq.~\eqref{eq:inacc}, and the minimal requirements 
$\vareps=0.5$~\cite{Damour:2010zb,MacDonald:2011ne} and $\vareps_{\rm
  M}=0.005$ or $\vareps_{\rm M}=0.035$~\cite{Lindblom:2008cm}. Note 
that they correspond to mismatches of $0.5~\%$ and $3.5~\%$,
respectively, where a $3.5~\%$ mismatch indicates that no more then
$10~\%$ of the signals are lost. 
Overall our results indicate that: 

(i)~the extrapolated waveform form the series HH$\{23456\}$ is effectual
and faithful for SNR $\vrho\lesssim10$ for most of the 
configurations if errors are computed from run HH6 and finite
extraction effects are neglected or included in both $h_{\rm x}$ and
$h_{\rm m}$ at the same time;  

(ii)~the extrapolated waveform HH$\{2346\}$ and the extrapolated
HH$\{23456\}^*$ with errors from HH6, are effectual only for the
less restrictive requirement, $\vareps_{\rm M}=0.035$, and faithful
for SNR $\vrho<3$; 

(iii)~if waveforms are extrapolated from less runs and/or computing
errors from runs at lower resolutions then HH5, the inaccuracy had
always larger or comparable values to those reported in the table for
HH$\{236\}$. 

In conclusion, minimal requirements for data analysis are met if waves
are extrapolated in resolution from more then four runs and certain
optimistic choices for the error bars are made. In the other cases
waveforms are inaccurate. Similar statements can be made if the
inaccuracy functional is computed up to $M\,\omega_{22}=0.1$, while
obviously its values increase slightly.

\section{Comparison with post-Newtonian T4 phasing formula}
\label{sec:comparepn}

In this section we perform a comparison between the NR
waveforms and PN approximants. The main goal is to quantify their 
agreement/disagreement and the relative signature of the tidal
interactions on the waves during the last nine orbits of the merger
process. 
The comparison presented here is not exhaustive;
a systematic investigation of the different phasing formulas (see
e.g.~\cite{Boyle:2007ft}) and fitting models, as well as the
investigation of different comparison procedures
(e.g.~\cite{Hannam:2010ec}), is beyond the scope of this work.  

Here we will focus only on the so-called T4
formula~\cite{Blanchet:2001ax,Damour:2002kr,Buonanno:2002ft,Blanchet:2004ek,Baker:2006mp,Boyle:2007ft},
T4pp hereafter, accurate at 3.5~PN level.
In addition to the point-particle T4, we will consider a ``tidal'' T4, T4td hereafter,
as proposed in~\cite{Hinderer:2009ca,Damour:2009wj,Vines:2010ca,Vines:2011ud,Baiotti:2011am}. 
T4td includes the leading-order (LO) and next-to-leading-order (NLO) tidal PN
corrections in the dynamics and the leading-order corrections in the
waveform~\footnote{ 
  With LO and  NLO we refer to tidal corrections
  of order $\O(x^5)$ and $\O(x^6)$ respectively in the PN expansion,
  where $x=(GM\Omega/c^3)^{2/3}$ is the PN expansion parameter.
}. 

A comparison with the T4 approximants and NR data have been already
considered in~\cite{Baiotti:2010xh,Baiotti:2011am},  
to which we also refer for the precise equations used in this work.
The analysis there is performed in frequency domain considering a
certain measure of the phase acceleration,  
that has the advantage of being independent on time and phase shifts
(and a simple physical interpretation, see discussion in Sec.~IV) 
but the drawback of requiring fits of the numerical data and a certain fine tuning.
The result obtained is that T4td NLO accumulates about $2.25$~$\rad$ on
the frequency interval $M~\omega_{22}\in[0.043, 0.057]$ for the model
employed here, and about $2\pi$~$\rad$ on the same interval for a binary
with less compact stars. 

In the following both T4pp and T4td are considered in a time-domain
comparison with the NR waveform.  
In order to be contrasted, the waveforms must be aligned in time and phase.
Let us make some general comments on this point.
There is no unique way to align waveforms for such comparison, in the
literature several methods are proposed,
e.g.~\cite{Boyle:2008ge,Read:2009yp,Hannam:2010ec}. A priori none of
them is free from ambiguities or clearly preferable.
The alignment region is typically chosen after the ``initial
transient'' (or adjustment) of the numerical waveforms. The transient
is related to the use of conformally flat initial data, and the 
main effect (but in principle not the only one) is the well-known
burst of radiation at early times of the simulation~\footnote{ 
  Note that in the plots in this paper the burst is not visible as a
  consequence of the integration algorithm to recover $h$ from
  $\psi_4$, but it is actually present in the $\psi_4$ data. The
  integration can also be performed in a way which takes into account 
  the burst~\cite{Damour:2011fu}.}. 
The transient is quite rapid, typically within the first
orbit, after that the system relaxes to the
expected quasicircular
state~\cite{Damour:2011fu}.
The allowed alignment region is
constrained by the validity of the post-Newtonian approximation and 
the length of NR waveforms. 
The lowest frequency interval, compatible with the NR data available
and the comment above, may be thus preferable.
For a quantitative analysis on the length requirement of NR waveforms
in the BBHs case see~\cite{Hannam:2010ky}.  

Guided by these considerations, we chose the following strategy:
(i) NR and PN waveforms are aligned in phase and time by considering
an interval, $[t_1,t_2]$, in the first half of the numerical signal available, where
the frequencies are closer to those of validity of the PN method;  
(ii) following~\cite{Boyle:2008ge} the time shift, $\Delta^s t$, and
the phase shift, $\Delta^s\Phi$, are determine by minimizing the
functional,
\be
G[\Delta^s t, \Delta^s\Phi]  =
\int_{t_1}^{t_2} dt\left[ \Phi(t)-\Phi^{\rm PN}(t-\Delta^s t)
-\Delta^s \Phi \right]^2 \ .
\ee
The PN waveforms are then matched to NR ones by applying the shifting;  
(iii) different results are obtained if the center and the length of the alignment interval are varied. 
However we observed that the main dependence is on the position of the
center rather than in the interval length. For simplicity, we fixed
the interval length as $100\,M$ and vary the position of the center
$t_c\in[0,900]\,M$;  
(iv) The best value of $t_c$ is estimated by minimizing the
mismatch between the PN and NR waveform. 

\begin{figure*}[t]
  \begin{center}
    \includegraphics[width=0.49\textwidth]{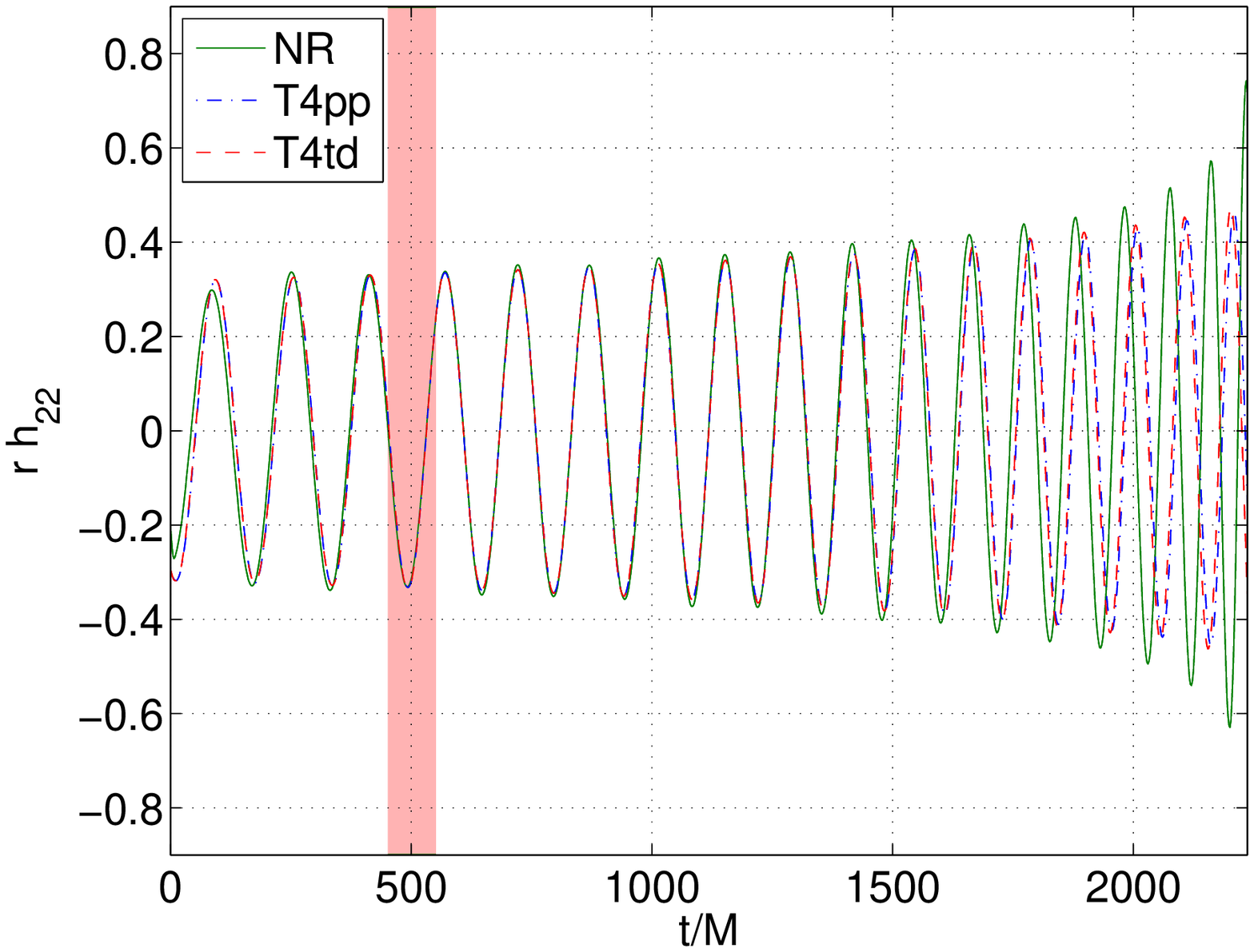}
    \includegraphics[width=0.49\textwidth]{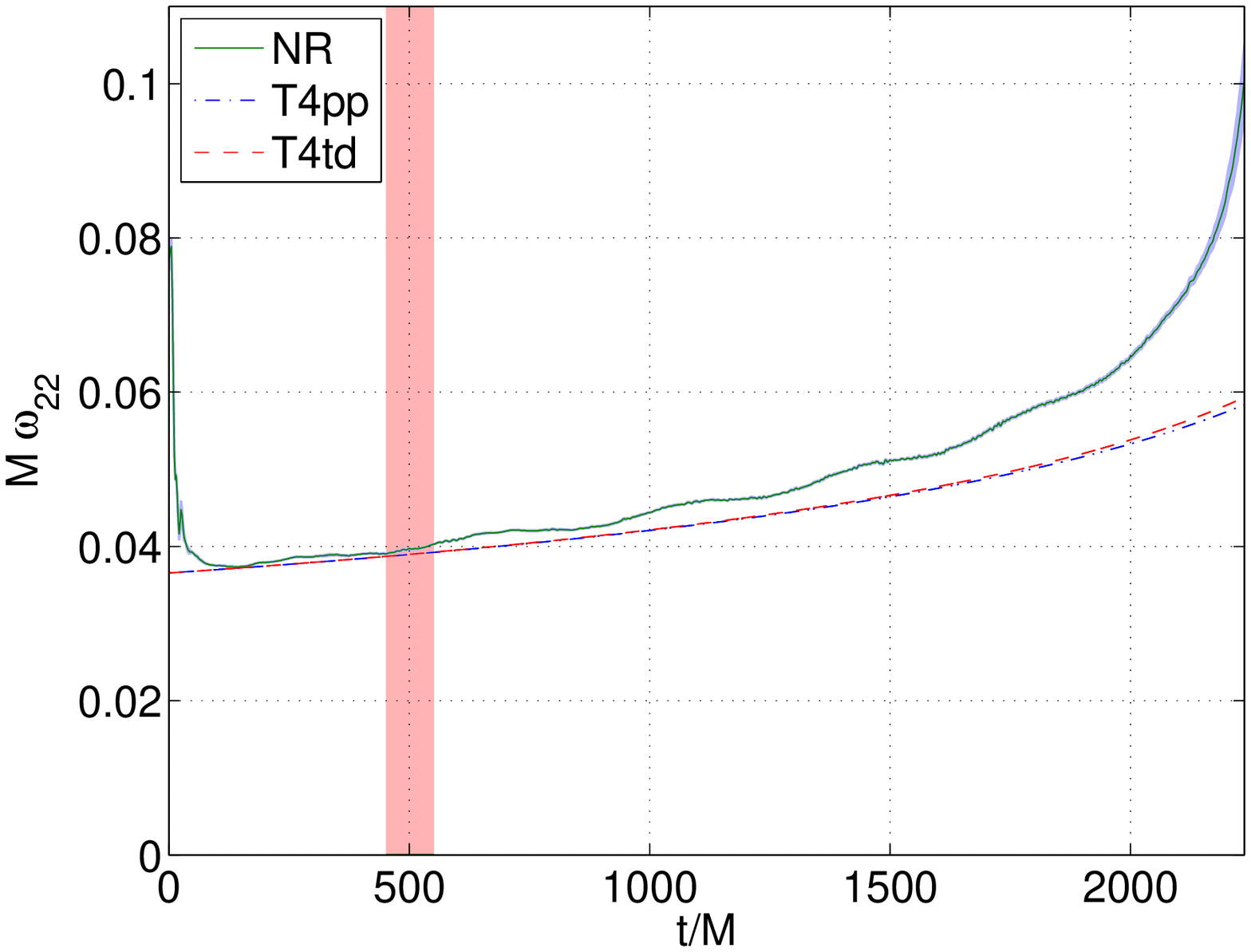}
    \caption{\label{fig:t4nr:h} Comparison T4pp/T4td--NR, 
      $r\,h_{22}$ (right) and $M\,\omega_{22}$ (left). 
      The shaded red area at early time is the alignment
      region.
      A very thin shaded blue area (barely distinguishable on this
      scale) shows the uncertainty of NR data.}  
  \end{center}
\end{figure*}

As a case study we focus on our best NR waveform extrapolated only in
resolution, i.e.~HH$\{23456\}$.
Neglecting finite-extraction uncertainties does not particularly
affect the conclusions, also because they decrease towards merger. 
As discussed extensively in~\cite{Hannam:2010ec} the error estimates
of Sec.~\ref{sec:waves} based on the convergence analysis may not be 
the optimal ones to be used in PN comparison. If only a certain
range of frequencies of the NR waveform is of interest, a phase error 
estimated on that range (i.e.~by shifting in some way NR waves from different runs)
may be less conservative and thus preferable for the specific application.
However, such error estimates suffer of ambiguities related to the
alignment procedure and we prefer not to pursue that method.
For the purpose of this section is sufficient and justified to use the
errors estimated in Sec.~\ref{sec:waves} that can be, eventually,
considered as an upper bound to the actual errors (see below).

\begin{figure}[t]
  \begin{center}    
    \includegraphics[width=0.49\textwidth]{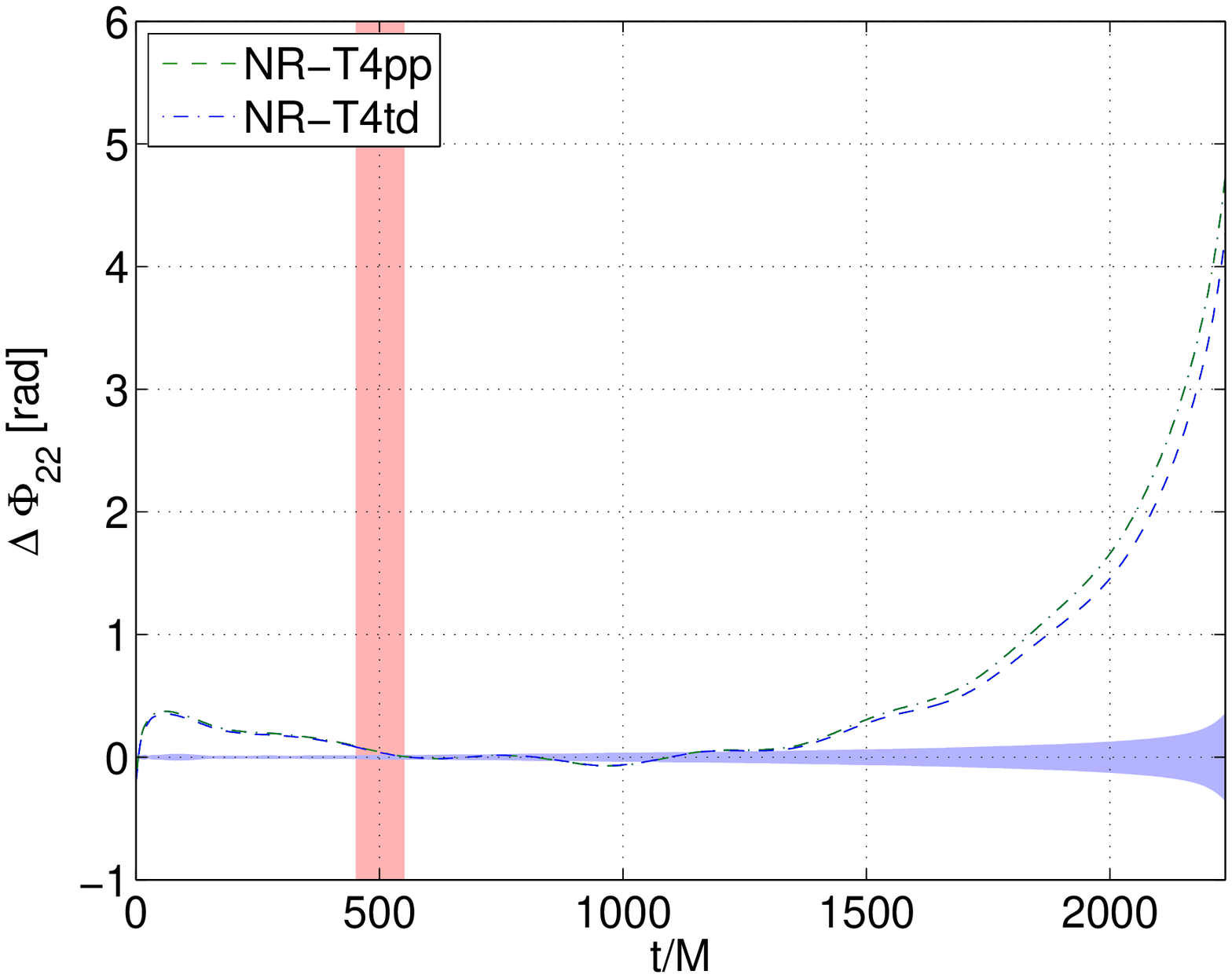}
    \includegraphics[width=0.49\textwidth]{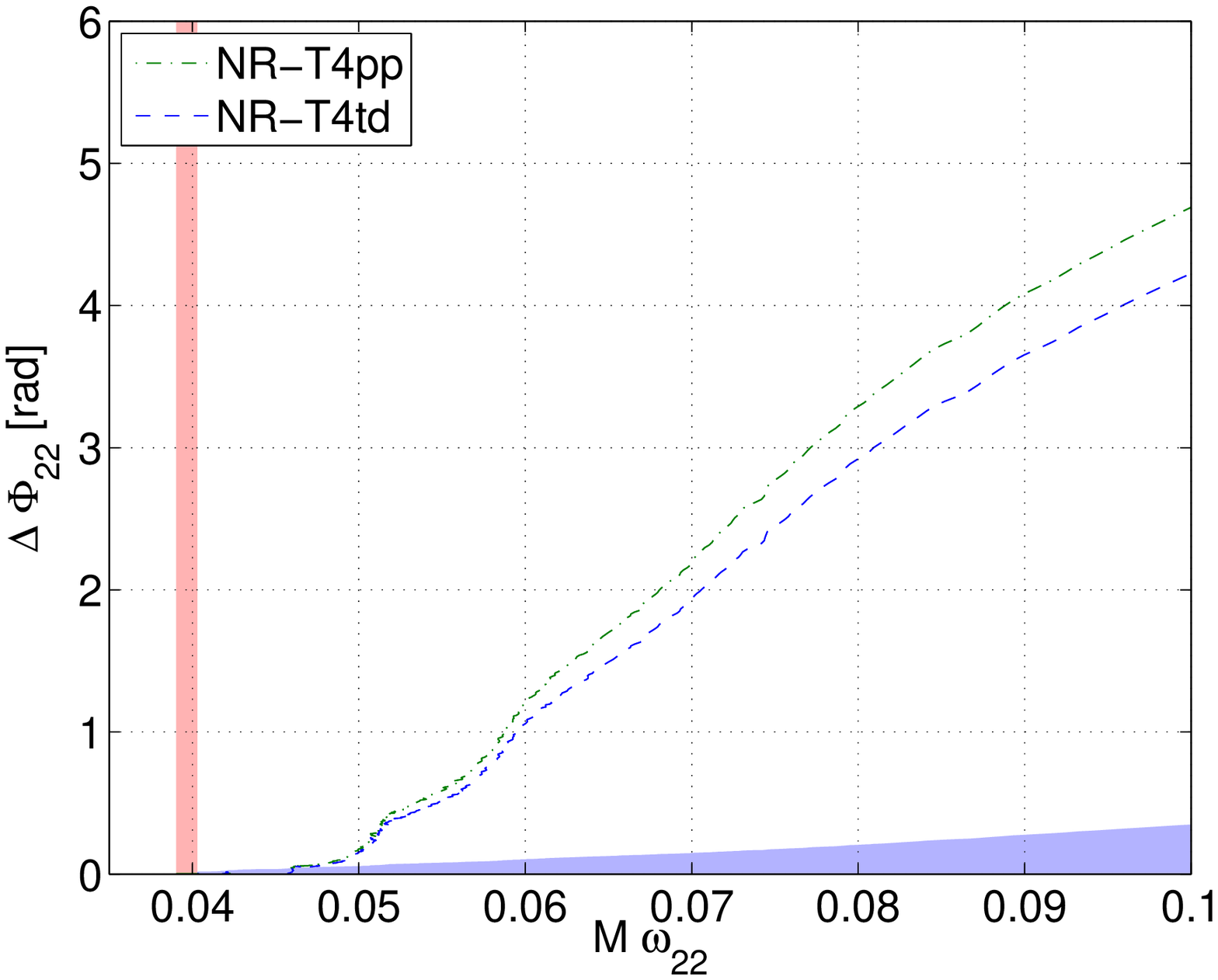}
    \caption{\label{fig:t4nr:phasee}
      Phase differences T4pp/T4td-NR
       in time (left) and frequency domain (right).
       The shaded red area at early time is the alignment region. 
       The shaded blue area is the uncertainty of NR data.}
  \end{center}
\end{figure}

Figure~\ref{fig:t4nr:h} shows the real part of the aligned waveforms
(error bars are not shown there for clarity), and the GW frequencies
with error bars. 
The alignment interval used for the analysis in the figure is
$[t_1,t_2]/M=[450,550]$, which minimizes the mismatch between the NR
and T4td waveform as described above. 
Figure~\ref{fig:t4nr:phasee} shows the phase differences between the
waveforms in time (left) and frequency 
(right) domain. The PN waveform maintains a good phasing for few GW
cycles after the alignment region and up to $t/M\sim1200$. 
At later times, phase differences with respect the PN evolution become
positive and significant, the largest difference is the one with T4pp.
At higher GW frequencies than $M\,\omega_{22}\sim0.05$ the PN
approximant significantly differ from NR waves, and the T4pp rapidly
accumulates a phase difference of 
$\Delta\Phi_{22}\sim 2$~$\rad$ at $M\,\omega_{22}\sim0.07$ and of
$\Delta\Phi_{22}\sim 4$~$\rad$ at $M\,\omega_{22}\sim0.1$.  
The T4td performs slightly better then T4pp at later
times, but (not yet calculated) higher order tidal corrections are
important~\footnote{ 
  We mention here that the LO tidal correction in the T4td waveform
  has a negligible effect: the main difference between T4pp and T4td
  is the correction in the PN dynamics.
  In addition, the performances of T4td may be strongly driven by the
  point particle part of the phasing formula, for which a resummed PN
  technique could be preferable~\cite{Damour:2009wj,Baiotti:2011am} (see
  also~\cite{Damour:2008te,Boyle:2008ge} for the BBH case).    
}. 
This is the central observation here: 
tidal interactions in the nonlinear regime dominate
the dynamics and the GW emission \emph{at least} during the last 5-6
orbits of the merger process. A similar conclusion can be drawn
considering the waveform HH$\{2346\}$, but not for HH$\{236\}$.
In the latter case the PN and NR signals are indistiguishable due to
larger error bars.

As mentioned above, we did not perform a systematic study of different
alignment procedures, but a certain dependence on the alignment
interval was expected and observed. The phase differences given above
are lower bounds, since they are determine by a minimization of the
mismatch functional. In particular, dropping the step (iv) in the
procedure outlined above and varying $t_c\in[0,900]\,M$, we estimated 
also an upper bound. The latter corresponds to an alignment interval
centered at $t_c\sim70\,M$ and the accumulated phase are
$\Delta\Phi_{22}\sim3$~$\rad$ at $M\,\omega_{22}\sim0.07$ and of
$\Delta\Phi_{22}\sim6$~$\rad$ at $M\,\omega_{22}\sim0.1$.   

\section{Conclusions}

In this paper we have presented results about the accuracy of NR
waveforms from BNS mergers and their comparison with PN methods. 
The simulations cover nine orbits of the late inspiral and the merger
phase, they are the longest and most accurate BNS simulations to date, 
in terms of the resolution employed and the number of runs performed for a single 
initial configuration.
The convergence of the waveforms and their uncertainties related to
truncation errors and finite-radius extraction are discussed.
For the first time in case of BNS merger 
waveforms, the accuracy standards for detection have been evaluated. 
The aim of the study is to assess the quality of NR waveforms  
in view of their future use to understand the physics of the merger
and of tidal interactions or for data analysis purposes. 
As a first step in this direction, a comparison with the PN T4
waveforms has been presented.

NR waveforms are found to be convergent at second order rate during the
inspiral and up to contact, i.e.~until the last 1.5 orbits or,
equivalently, for the GW frequencies $M\,\omega_{22}\lesssim0.07$. 
Later an over-convergent behavior is observed, likely 
due to the numerical treatment of the matter and possibly also to the
lowest resolution run employed in the self-convergence test.
The uncertainties on the inspiral waves have been estimated
by using Richardson extrapolation of different data sets and
assuming the observed convergence rate where it is valid. 
finite-radius extraction affects were investigated by extrapolating
the waves to null-infinity. 
Truncation errors increase towards the merger, when the amplitude of
the GW become maximum. The maximum errors in phase and amplitude
observed are reported in Tab.~\ref{tab:maxerrs} for different
extrapolations series. The errors related to finite-radius extraction
decrease towards the merger and they are generically the smaller then
truncation errors, but of the same order of magnitude of truncation
errors in some relevant cases of waves extrapolated in resolution.     
Some of these results are compatible with the findings
of~\cite{Baiotti:2011am}.  

Accuracy standards have been evaluated using noise curves of
ground-based detectors, and assuming minimal requirements. 
Results are reported in Tab.~\ref{tab:acc}.
Considering the most optimistic error-bars, extrapolated waveforms from
five runs are effectual and faithful for detection 
with SNR $\vrho\lesssim10$ for most of the configurations considered. 
Considering instead more conservative error-bars, or extrapolation in
resolution with fewer runs, waveforms are neither effectual nor 
faithful for relevant SNRs. 

These facts may affect some of the conclusions of previous works 
where errors of the NR waveforms were (not available and thus)
neglected, but statements about the detectability of (small)
effects (EoS, magnetic fields) were made.
Our results should be taken into account and/or reproduced in future
works employing NR waveforms for data analysis purposes or for
comparison with analytic methods.  

The NR data have been compared with the prediction of the PN T4 formula,
both for point-particle (T4pp) and including all the analytically
known tidal corrections (T4td). 
The comparison between the NR and the T4 waves has been
carried out by aligning them in time and phase at low frequencies, and
looking at the accumulated phase difference. 
The aligned T4pp waveform accumulates rapidly a significant dephasing of
$\Delta\phi_{22}\sim2$ 
at $M\,\omega_{22}\sim0.07$, 
and $\Delta\phi_{22}\sim4$ 
at $M\,\omega_{22}\sim0.1$. 
These values can be considered lower bounds since in the comparison waves 
are aligned in such a way to minimize the mismatch and error-bars from 
the convergence test are employed.
The inclusion of tidal corrections does not reduce the phase difference 
more than a fraction of a radiant.
The results suggest that tidal interactions are very amplified in a strong field and
nonlinear regime and play a significant role already during the last nine
orbits. As already observed in~\cite{Baiotti:2011am} the analytically
known LO and NLO tidal terms in the T4 PN approximant are not sufficient to
match the NR waveform. 

In summary, our work indicates that NR waveforms from BNS are
physically ``reliable'' because convergent and comparable with PN at
sufficiently low frequencies. 
The measured uncertainties are such that the NR waveforms from BNS 
may not be sufficiently accurate for data analysis purposes, 
unless data extrapolated from several runs are employed.
For data analysis applications, an error estimate based on relative
as well as absolute comparisons (aligning/not-aligning the waveforms) will
be relevant.
A very careful evaluation of the waveform uncertainties is unavoidable for
their use in quantitative studies. 

Future work will be devoted to a more comprehensive comparison between
the NR waves and analytic PN fitting models, either in
the standard Taylor form or resummed one~(EOB), to extend these
results to different mass ratios and EoSs, and to the investigation of
different grid setup and higher order methods for the treatment of the matter.

\appendix

\section{Accuracy standards}
\label{app:astd}

In this appendix the accuracy standards used in this paper are
discussed. We
follow~\cite{Cutler:1994ys,Miller:2005qu,Damour:2010zb,Lindblom:2008cm,Lindblom:2009ux}.  

Given two real time series (waveforms), $h_{\rm x,m}(t)$, and their 
Fourier complex transform, $\tilde{h}_{\rm x,m}(f)$, the Wiener scalar product is
defined as, 
\be
\label{eq:overlapint}
(h_{\rm x}|h_{\rm m}) \equiv 4 \Re \int_0^\infty df 
\frac{\tilde{h}_{\rm x}(f)\tilde{h}^*_{\rm m}(f)}{S_{\rm n}(f)}\ ,
\ee
where $S_{\rm n}(f)$ is the one-sided power spectral density of the
detector noise. The (squared of the) norm associated to the Wiener
product is $||h_{\rm x}||^2_{\rm w} \equiv (h_{\rm x}|h_{\rm x})$.
The Wiener product is real and symmetric, the associated norm is
positive definite. The Wiener product provides a measure in the
waveform space~\cite{Cutler:1994ys}. The \emph{mismatch functional} is
defined as, 
\be
\M[h_{\rm m},h_{\rm x}] \equiv 1 
- \frac{(h_{\rm x}|h_{\rm m})}{||h_{\rm x}||_{\rm w} ||h_{\rm m}||_{\rm w}} ,
\ee
and it is often employed to discuss accuracy standards for BBH NR
waveforms, see e.g.~\cite{Hannam:2009hh,Reisswig:2009vc,Hannam:2010ky}.
In this work we will mainly focus on (the square of) the
\emph{inaccuracy functional}~\cite{Damour:2010zb},  defined as
\be
\label{eq:I2}
\mathcal{I}^2[h_{\rm m},h_{\rm x}] = 
\frac{|| h_{\rm m}-h_{\rm x}||_{\rm w}}{||h_{\rm x}||_{\rm w}} \ .
\ee
The inaccuracy functional has been considered for NR waveforms, for
example, in~\cite{MacDonald:2011ne}, and provides an equivalent measure
to the mismatch functional.

Assuming an ideal detector (i.e.~neglecting calibration errors), 
a waveform $h_{\rm m}$ (the ``model'') is
indistinguishable from the waveform $h_{\rm x}$ (the ``exact''),
if and only if their difference $\delta h = h_{\rm m}-h_{\rm x}$ satisfies,
\be
\label{eq:faith}
||\delta h||_{\rm w} < 1 \ .
\ee
Eq.~\eqref{eq:faith} represents an accuracy requirement for
measurement purposes, and it determines the \emph{faithfulness} of the
model waveform.  
A less restrictive requirement can be given for detection purposes,
and it is related to the \emph{effectualness} of the model waveform. 
A sufficient condition is, 
\be
\label{eq:effect}
||\delta h||_{\rm w} < \sqrt{2\vareps_{\rm M}} \vrho \ ,
\ee
where $\vrho\equiv ||h_{\rm x}||_{\rm w}$ is the
optimal signal-to-noise ratio (SNR) and the constant
$\vareps_{\rm M}$ set the accuracy level. 
We set in this work $\vareps_{\rm M}=0.005$ or $\vareps_{\rm M}=0.035$
as suggested in~\cite{Lindblom:2008cm} (see references therein).

Conditions~\eqref{eq:faith} and~\eqref{eq:effect} depend on the
distance of the detector form the source. If they are written in term
of the inaccuracy functional, the dependence on absolute scales can be
moved to the right-hand-side and expressed only in term of the
SNR. The accuracy requirements then read, 
\be
\label{eq:inacc}
\mathcal{I}^2 = \frac{||\delta h ||_{\rm w}}{||h_{\rm  x}||_{\rm w}}
< \begin{cases}
\vareps /\vrho          & \text{faithful} ,\\
\sqrt{2\vareps_{\rm M}}  & \text{effectual} , \\
\end{cases}
\ee
where the functional dependence has been omitted for clarity. 
The level $\vareps<1$ is here set as $\vareps=0.5$.  
Equivalent conditions to Eq.~\eqref{eq:inacc} can be expressed in term
of the L2 norm of the time domain signals~\cite{Lindblom:2008cm},
however they are not considered here because they seem more
restrictive than the frequency domain criteria.

In order to evaluate the accuracy of NR waveforms, one can consider
$h_{\rm x}$ as the best waveform model 
(the extrapolated in resolution in our case), 
and construct $h_{\rm m}=h_{\rm x}+\delta h$ from the error estimates 
(the last resolution waveform in our case).
The accuracy requirements in Eq.~\eqref{eq:inacc} then quantify the
accuracy of the NR waveforms.
In the analysis presented in the paper the integration interval
$f\in[0,\infty]$ is approximated as $f\in[f_0,f_{\rm max}]$, covering
only the inspiral physical frequencies. 

The Wiener product is computed by scaling the waveforms to
physical units and to an effective distance, ${\rm D_{eff}}$, typically
given in $100~Mpc$. From the code output $r\,h_{22}$, we: 
(i)~recover $r\,h_+$ from the (2,2) multipole, using the expression
for the spin weighted spherical harmonics, 
${}^{-2}Y_{2\pm2}(\theta,\phi)=\sqrt{5/(64\pi)}\exp(\pm
i2\phi)(1\pm\cos\theta)^2$;  and 
(ii)~scale $r\,h_+$ to an effective distance.
Assuming the radiation is emitted on the $z$ axis, perpendicularly
to the orbital plane, one has,
\bea
r h_{+}(t) &=& r\ \Re\left( {}^{-2}Y_{22} h_{22} + {}^{-2}Y_{2-2}
h_{2-2} \right) \\
&\simeq& 0.6308 \ r\ \Re\left( h_{22}\right) \ \ \ (\mbox{for}\ \theta=0\, ,\ \phi=0 )\non\\
h_{+}(t,D_{\rm eff}) &=& r h_{+}(t) \ {\GMc2} \ \left(\frac{{\rm D_{eff}}}{\Mpc}\right)^{-1}\\
&\simeq& r h_{+}(t) \ 4.7857 \times 10^{-20} \ \left(\frac{{\rm
    D_{eff}}}{\Mpc}\right)^{-1} \ . \non
\eea
Unit conversion: $1~\Mpc\simeq3.08568025 \times 10^{24}$~$cm$, $\GMc2\simeq1.47670133 \times 10^5$~$cm$.

\begin{acknowledgments}
  The authors thank Mark Hannam, David Hilditch, and Alessandro Nagar 
  for discussions and reading the manuscript. 
  The authors thank Jocelyn Read for discussions about the PN comparison.
  The authors thank the Meudon group for making publicly available 
  {\lorene} initial data and Eric Gourgoulhon for explanations. 
  
  This work was supported in part by  
  DFG grant SFB/Transregio~7 ``Gravitational Wave Astronomy''.
  Computations where performed mainly on JUROPA (JSC, J\"ulich)
  and also at LRZ (Munich).
\end{acknowledgments}

\bibliographystyle{apsrev}     
\bibliography{refs20120522}{}

\end{document}